\documentclass[prb,preprint,amsmath,amssymb,superscriptaddress,notitlepage]{revtex4-1}

\usepackage{epsfig}
\usepackage{graphicx}
\usepackage{color}
\usepackage{bm}
\usepackage{fixme}
\usepackage{amsmath}
\usepackage{bm}

\begin{document}

\title{Quenching of an antiferromagnet into high resistivity states using electrical or ultrashort optical pulses}
\author{Z.~Ka\v{s}par}
\affiliation{Institute of Physics, Czech Academy of Sciences, Cukrovarnick\'a 10, 162 00, Praha 6, Czech Republic}
\affiliation{Faculty of Mathematics and Physics, Charles University in Prague, Ke Karlovu 3, 121 16 Prague 2, Czech Republic}
\author{M.~Sur\'ynek}
\affiliation{Faculty of Mathematics and Physics, Charles University in Prague, Ke Karlovu 3, 121 16 Prague 2, Czech Republic}
\author{J.~Zub\'a\v{c}}
\affiliation{Institute of Physics, Czech Academy of Sciences, Cukrovarnick\'a 10, 162 00, Praha 6, Czech Republic}
\affiliation{Faculty of Mathematics and Physics, Charles University in Prague, Ke Karlovu 3, 121 16 Prague 2, Czech Republic}
\author{F. Krizek}
\affiliation{Institute of Physics, Czech Academy of Sciences, Cukrovarnick\'a 10, 162 00, Praha 6, Czech Republic}
\author{V.~Nov\'ak} 
\affiliation{Institute of Physics, Czech Academy of Sciences, Cukrovarnick\'a 10, 162 00, Praha 6, Czech Republic}
\author{R.~P.~Campion}
\affiliation{School of Physics and Astronomy, University of Nottingham, Nottingham NG7 2RD, United Kingdom}
\author{M.~S.~W\"{o}rnle}
\affiliation{Department of Materials, ETH Zurich, 8093 Zurich, Switzerland}
\affiliation{Department of Physics, ETH Zurich, 8093 Zurich, Switzerland}
\author{P.~Gambardella}
\affiliation{Department of Materials, ETH Zurich, 8093 Zurich, Switzerland}
\author{X.~Marti}
\affiliation{Institute of Physics, Czech Academy of Sciences, Cukrovarnick\'a 10, 162 00, Praha 6, Czech Republic} 
\affiliation{IGS Research, Calle La Coma, Nave 8, La Pobla de Mafumet, Tarragona 43140, Spain}
\author{P.~N\v{e}mec}
\affiliation{Faculty of Mathematics and Physics, Charles University in Prague, Ke Karlovu 3, 121 16 Prague 2, Czech Republic}
\author{K.~W.~Edmonds}
\affiliation{School of Physics and Astronomy, University of Nottingham, Nottingham NG7 2RD, United Kingdom}
\author{S.~Reimers}
\affiliation{School of Physics and Astronomy, University of Nottingham, Nottingham NG7 2RD, United Kingdom}
\affiliation{Diamond Light Source, Chilton, Didcot, UK}
\author{O.~J.~Amin}
\affiliation{School of Physics and Astronomy, University of Nottingham, Nottingham NG7 2RD, United Kingdom}
\author{F.~Maccherozzi}
\affiliation{Diamond Light Source, Chilton, Didcot, UK}
\author{S.~S.~Dhesi}
\affiliation{Diamond Light Source, Chilton, Didcot, UK}
\author{P.~Wadley}
\affiliation{School of Physics and Astronomy, University of Nottingham, Nottingham NG7 2RD, United Kingdom}
\author{J.~Wunderlich}
\affiliation{Institute of Physics, Czech Academy of Sciences, Cukrovarnick\'a 10, 162 00, Praha 6, Czech Republic}
\affiliation{Institute of Experimental and Applied Physics, University of  Regensburg, Universitätsstrasse 31, 93051 Regensburg}
\author{K.~Olejn\'ik} 
\affiliation{Institute of Physics, Czech Academy of Sciences, Cukrovarnick\'a 10, 162 00, Praha 6, Czech Republic}
\author{T.~Jungwirth}
\affiliation{Institute of Physics, Czech Academy of Sciences, Cukrovarnick\'a 10, 162 00, Praha 6, Czech Republic} 
\affiliation{School of Physics and Astronomy, University of Nottingham, Nottingham NG7 2RD, United Kingdom}

\maketitle

{\bf 
Ultra-fast dynamics, insensitivity to external magnetic fields, or absence of magnetic stray fields are examples of properties that make antiferromagnets of potential use in the development of spintronic devices. Similar to their ferromagnetic counterparts, antiferromagnets can store information in the orientations of the collective magnetic order vector. However, also in analogy to ferromagnets, the readout magnetoresistivity signals in simple antiferromagnetic films have been weak and the extension of the electrical reorientation mechanism to optics has not been achieved. Here we report reversible and reproducible quenching of an antiferromagnetic CuMnAs film by either electrical or ultrashort optical pulses into nano-fragmented domain states. The resulting resistivity changes approach 20\% at room temperature, which is comparable to the giant magnetoresistance ratios in ferromagnetic multilayers. We also obtain a signal readout by optical reflectivity. The analog time-dependent switching and relaxation characteristics of our devices can mimic functionality of spiking neural network components.  
}

Ferromagnets and antiferromagnets are two fundamental classes of materials with magnetic order. Although antiferromagnets are more readily available than ferromagnets, their magnetic order cannot be easily harnessed \cite{Neel1971}. Current-induced spin-orbit fields in ferromagnets with broken inversion symmetry can be used to manipulate the magnetic order vector of a ferromagnet, an effect necessary for the operation of ferromagnetic memories. The reorientation of the magnetic order vector -- or switching -- has recently been extended to antiferromagnets such as CuMnAs and Mn$_2$Au \cite{Shick2010,Park2011,Chen2014,Nakatsuji2015,Zelezny2014,Wadley2016}, where the local symmetry-breaking of the lattice gives rise to a staggered spin-orbit field \cite{Zelezny2018,Wadley2016,Zelezny2014,Bodnar2018}, prompting further research into the symmetry and topology landscape of antiferromagnets \cite{Jungwirth2016,Baltz2018,Jungwirth2018,Zelezny2018,Nemec2018,Smejkal2017b,Gomonay2018}. In these first antiferromagnetic memories, the readout was based on the anisotropic magnetoresistance (AMR) and the retention originated from the switching between ground states with different directions of the collective N\'eel vector, similar to the operation of the first generation of ferromagnetic random access memories (MRAMs)  \cite{Daughton1992}.

The polarity of current pulses with pulse lengths in the $\mu$s--ms range and threshold currents around $10^6$~Acm$^{-2}$ was used to demonstrate the reorientation (switching) of the N\'eel vector in CuMnAs devices \cite{Wadley2018}. The observations, obtain by X-ray magnetic linear dichroism photoemission electron microscopy (XMLD-PEEM), showed reproducible, reversible, and stable switching of large, micron-scale antiferromagnetic domains and the measurements were consistent with the expected symmetry of the staggered current-induced field. However, only weak AMR readout signals of $\sim0.1$~\% were obtained by switching the large domains in the CuMnAs film \cite{Wadley2018}. The small AMR amplitude in single-layer CuMnAs films was independently confirmed in a recent study where the N\'eel vector reorientation was induced by a strong external magnetic field  \cite{Wang2020}. On the other hand, multilayers with giant-magnetoresistance, which made ferromagnetic MRAMs a commercially viable technology \cite{Chappert2007}, have not been realized in antiferromagnets.

Furthermore, after a decade of research, optical control of magnetism using ultra-short single femtosecond-laser pulses has been demonstrated only for ferrimagnets \cite{Kimel2019}; the reorientation of the ferrimagnetic order vector is based on unequal, light-induced demagnetization dynamics of the different spin-sublattices and this mechanism is not compatible with antiferromagnets. All-optical switching was recently observed in ferromagnets, but requires a complex process involving demagnetization and domain-nucleation and expansion driven by multiple laser pulses  \cite{Kimel2019}.

In this Article, we report deterministic and reversible changes in the resistivity of a film of antiferromagnetic CuMnAs by using electrical pulses with a current density over $10^7$~Acm$^{-2}$ or by ultra-short optical pulses.  By applying the writing pulses, the antiferromagnet is brought to the vicinity of the magnetic transition point and then quenched into a nano-fragmented domain state with high resistivity. We term this effect quench switching of an antiferromagnet. The corresponding resistivity changes exceed the previously reported AMR signals related to the reorientation switching of the magnetic order vector by up to three orders of magnitude. The quench switching can by also probed optically by reflectivity measurements. The nano-fragmented antiferromagnetic domain state under electrical pulsing conditions has been imaged by diamond NV-center magnetometry  \cite{Wornle2019}, and we confirm the findings  via synchrotron-based XMLD-PEEM measurements.  Our observations suggest that the time dependence of the resistivity after the quench switching  follows the Kohlrausch stretched-exponential relaxation form.

\newpage
\noindent{\large\bf Electrical switching detected by resistivity or reflectivity}

\noindent Fig.~1 shows results of the quench switching by unipolar electrical current pulses in a bar resistor at room temperature. Our microdevices (Fig.~1b) were lithographically patterned from a 50~nm thick epilayer of tetragonal CuMnAs of 20~$\Omega$ sheet resistance grown by molecular beam epitaxy on an insulating lattice-matched GaP substrate (Fig.~1a) \cite{Wadley2013,Krizek2020}. In the measurements plotted in Fig.~1c, we use the simplest two-point resistor geometry for delivering the writing pulses and for detecting the readout signals. We start by applying an electrical writing pulse of 100~$\mu$s and measure for 1~s the resulting change in the ohmic resistance of the device. The pulse induces an increase  in the resistance. We then repeat the measurement using a writing pulse of reduced amplitude (by 8\%) and observe that the resistance drops back to a lower value. This pattern of changes in the resistance is precisely reproducible as illustrated after ten pulse sequences, corresponding to high/low resistance levels (Fig.~1c). A different pulsing sequence confirming the reproducible and deterministic characteristics of the process is shown in Fig.~1d (see also Supplementary Information Fig.~S1).

The abrupt resistance drop after the weaker pulse excludes the possibility that the detected high resistance after the first pulse was due to the sample still remaining at an elevated temperature during the readout time. In the Supplementary Information Fig.~S2 we show that within a few $\mu$s after the pulse, the sample cools down back to the base temperature. Our switching  signals are therefore detected at times safely exceeding the transient heating time. We estimated from the measured resistivity during the 100~$\mu$s pulse  and from the measured dependence of the resistivity on the base temperature (Supplementary Information Fig.~S2) that the temperature of the device during the pulse reaches approximately $200\pm 50^\circ$C, i.e., is around  the CuMnAs  N\'eel temperature  \cite{Wadley2013}.  

In Fig.~2a we plot the relative size of the quench-switching signal, $\delta\rho/\rho$, obtained directly from a four-point resistivity measurement (see dashed voltage probes in Fig.~1b).  The data are plotted for amplitudes of the writing pulses safely below any detectable parasitic electromigration effects (see the discussion of structural characterization below and in the Supplementary Information Sec.~II).  In all electrical pulsing experiments we use a voltage pulse-source. The voltage-current characteristics for the 100~$\mu$s writing pulse is shown in Supplementary Information Fig.~S2. In Fig.~2a we see no switching signal at lower current amplitudes and an onset followed by an increase with increasing pulse amplitude \cite{Shi2020}. In all measurements, the reading current amplitude is well below the onset writing current. The resistive switching ratio approaches 20\% at room temperature, i.e., is comparable to the giant-magnetoresistance ratios in ferromagnetic multilayers \cite{Chappert2007}.

While at 300~K the metastable high-resistivity state relaxes within a $\sim10$~s time-scale, at 200~K the retention time approaches a year (for details see the discussion of Fig.~4). When quench-switching the device by the 100~$\mu$s writing pulse at 200~K we can therefore perform longer measurements, such as the temperature sweep shown in Fig.~2b,  with no relaxation of the written high-resistivity state. By comparing the temperature dependence of the resistivity after applying the writing pulse, $\rho+\delta\rho$, to the resistivity without the pulse, $\rho$, we see that the switching signal $\delta\rho$ is nearly temperature independent while $\rho$ decreases with decreasing temperature in the metallic CuMnAs. As a result,  the switching ratio approaches 100\% at low temperatures.  We point out that the simple bar-geometry of our devices and their micron-scale lateral size were chosen to highlight the bulk resistivity nature of our quench-switching signals. A comparison of CuMnAs films with thicknesses between 50 and 20~nm further verifies that the signal stems from the bulk of the antiferromagnetic metallic film and not from interface or surface effects. 

In Fig.~3  we show measurements on a Wheatstone-bridge microdevice which brings several technical advantages over the simple bar geometry. By pulsing the bridge first along one pair and then along the orthogonal pair of resistors and by measuring the bridge resistance $R_{\rm T}$ (Fig.~3a) we can directly check the reproducibility of the switching signal  between different physical resistors while using the same readout probes. Apart from the flipped sign of $R_{\rm T}$, which is a geometry effect of the Wheatstone-bridge, the two pairs of resistors show a quantitatively identical switching pattern, as seen in Fig.~3b.  This includes the increase of the switching signal with the number of successive pulses of the same amplitude, and the relaxation after each pulse (for more examples see Supplementary Information Figs.~S3-7). 

Figs.~3c,d further highlight the analog time-dependent characteristics of our switching signals. When the successive writing pulses have different amplitudes, the written signal can encode the order and delay between the pulses. The presence of a falling edge indicates that the weaker pulse arrived after the stronger pulse and the size of the falling edge decreases with increasing the delay between the two pulses. Supplementary Information Figs.~S8,9 highlight the reproducibility and show a more systematic mapping of this analog time-dependent characteristics. We point out that highly reproducible switching characteristics are obtained for different CuMnAs wafers prepared under equivalent growth conditions, and similar switching behavior is also seen in films grown on GaAs or Si over a range of growth parameters\cite{Krizek2020}, and in different device geometries (cf. Fig.~1, Fig.~S1, Fig.~3, Fig~S3-10).

Another advantage of the Wheatstone-bridge geometry is that it removes the temperature dependent offset resistance. (Without applying the switching pulse, the bridge is balanced and $R_{\rm T}=0$.) This allows us to perform low-noise measurements of the switching and subsequent relaxation over times spanning more than four orders of magnitude up to timescales of hours, as shown in Fig.~4a. The observed functional form of the relaxation is universal to all studied temperatures and contains two leading components accurately fitted to Kohlrausch stretched exponentials, $\sim\exp[-(t/\tau_{1(2)})^{\beta}]$, with $\beta=0.6$ and two different relaxation times. In theories of relaxation in complex systems, such as glassy materials, $\beta=d/(d+2)$ where $d$ is the dimensionality of the diffusion equation describing the relaxation process \cite{Phillips2006}. 
Our experimental value of $\beta$ corresponds to $d=3$ in the measured 50~nm thick film. 

Consistent with theory \cite{Phillips2006}, the dependence of the relaxation times on temperature follows a simple exponential, $\tau_{1(2)}=\tau_0\exp(E_{1(2)}/k_BT)$, as seen in Fig.~4b. Here the extrapolated attempt time $\tau_0$ is within the picosecond range typical of antiferromagnetic dynamics, compared to ferromagnets where it falls into the ns range \cite{Aharoni1969}. For the activation energy we obtained $E_1/k_B=30.8\times 300$~K for the slower component and $E_2/k_B=26.1\times 300$~K for the faster component, giving the room-temperature relaxation times in the $\sim10$~s and $\sim10$~ms ranges, respectively.  (For more details on the fitting see Supplementary Information Fig.~S11.)  We point out that the exponential temperature dependence of the relaxation time explains the unipolar switching from the high to low resistivity state  seen in Figs.~1 and 3. During the transient heating of the device due to the weaker pulse, the relaxation of the signal written by the preceding stronger pulse is exponentially accelerated which results in the abrupt falling edge of the signal after the weaker pulse. The resulting signal written by the weaker pulse is lower which reflects the dependence of the size of the signal on the amplitude of the writing pulse (cf. Fig.~2a). 

The picture of switching in our devices into complex states, as described by the Kohlrausch model,  is consistent with the antiferromagnetic domain nano-fragmentation systematically imaged in our devices by laboratory diamond magnetometry\cite{Wornle2019} and confirmed by our synchrotron X-ray microscopy which we discuss later in the article. The Kohlrausch relaxation is an example, noted in the past in a range of physical systems, that reproducible and universal  behavior can transcend the details of the complex disordered materials and of the measurement probes\cite{Phillips2006}. Indeed, we observe high reproducibility of the above switching and relaxation characteristics across different samples and experimental set-ups.

In Figs.~4c,d we demonstrate that our current-induced switching can be probed not only electrically but also optically by measuring the reflectivity change.  We again apply the stronger/weaker electrical writing pulse sequence and detect the falling edge. In this experiment we set the delay between the pulses to 5~ms to further illustrate the dynamics of the faster relaxing component   ($\tau_2$) of our switching signal. The short measurement window also allowed us to minimize the noise due to the optical set-up instability in the detection of the weak change in the optical reflectivity. Apart from the smaller amplitude, the optical signal (Fig.~4c) shows the same switching and relaxation pattern as the electrically detected signal (Fig.~4d). 

For consistency, we used 100~$\mu$s writing pulses in all  experiments presented so far. Our devices have, however, analogous switching characteristics over the full explored range of current pulse lengths down to 1~ns. This is illustrated in Fig.~4e, again on the stronger/weaker writing pulse sequence and the detection of the falling edge (for more results see Supplementary Information Fig.~S12.).  We point out that the writing current density  in Fig.~4e  is lower than in the spin-orbit torque switching by ns-pulses in conventional ferromagnetic devices  \cite{Baumgartner2017}.  

\bigskip
\noindent{\large\bf fs-laser pulse switching}

\noindent
To explore the ultimate pulse-length limit  we now move in Fig.~5 to the optical switching.  In the  measurement shown in Fig.~5b we used a $800$~nm wavelength laser beam with a 1.7~$\mu$m spot size and an energy of 1~nJ (30\% absorbed) per 100~fs pulse. Using a pulse-picker we select a single 100~fs pulse focused on a single spot on the Wheatstone bridge device (see Fig.~5a). 550~ms after the laser pulse we start recording the electrical readout  signal $R_{\rm T}$. We then shift the laser spot to the neighboring arm of the Wheatstone bridge and again perform the single-pulse switching measurement. In analogy to the electrical switching, the sign of $R_{\rm T}$ flips when shifting the spot from one to the neighboring arm and we observe the same relaxation characteristics of the signal.  The energy density required for switching by the single fs-laser pulse ($\sim$kJcm$^{-3}$) is comparable to the Joule energy density in the electrical switching by a ns-pulse (see Supplementary Information Fig.~S12). Since switching is governed by the energy (current) density, our  mechanism allows for the device scalability as in the latest generation of ferromagnetic spin-torque memories, and in contrast to the unscalable traditional magnetic recording by the Oersted  field proportional to current. We emphasize that our quench-switching mechanism allows us to write, in the same material and device,  by a pulse whose length covers the entire range from micro to femtoseconds and that our energy of the laser pulse is lower than in the earlier ultra-fast optical switching experiments in ferrimagnets\cite{Kimel2019}.

Next we show that we can optically control the high/low resistance switching as in the unipolar electrical experiment.  In the measurement in Fig.~5c we fixed the position of the laser spot in one arm and first applied the 1.7~$\mu$m focused laser beam with  10~kHz repetition rate over 250~ms. Then we measured for 5~s the readout signal. Next we applied another  pulse train in which we defocused the beam to a $2\times$ larger diameter to reduce its maximum energy. We observe the falling edge in the readout signal. In analogy to the electrical switching experiment in Fig.~3, the size of the falling edge depends on the delay between the stronger and the weaker laser pulse, as shown in Figs.~5c,d. 

The optical switching we observe is independent of the polarization of the laser pulse, as shown in Fig.~6a. The comparison between Figs.~5b,c also shows that the switching signals after a train of pulses are larger than after a single laser-pulse. The dependence on the number and amplitude of the laser pulses is studied in detail in Figs.~6b,c. Here to facilitate a direct comparison to the amplitudes in the electrical switching measurements, we show the  readout signal measured  after completing a  scan of the laser beam over the entire arm of the Wheatstone-bridge device. Using a pulse-picker, the points in Fig.~6b are obtained with a 600~ms long scanning pulse train while the repetition rate is increased from 10~kHz to 0.5~MHz. The switching signals generated by the optical pulses reach comparably high amplitudes as in the case of electrical pulses. Also in analogy to the electrical writing, we observe a multi-level optical switching but due to the high repetition rates of the laser pulses we can access orders of magnitude higher number of pulses. A single device is exposed to millions of pulses in these optical measurements which, apart from the reproducibility, evidences the endurance of the devices in our switching experiments.

To illustrate the additional spatial resolution allowed by extending our switching mechanism to optical pulses, we plot in Fig.~6d a scanning pixel-by-pixel sequence of optical-writing and electrical-readout in the Wheatstone bridge device. Here each micron-size pixel is exposed to a train of pulses with a 10~kHz repetition rate over 250~ms. Then  we record the electrical readout  signal $R_{\rm T}$ and shift the laser spot to the next pixel and repeat the same measurement procedure. As expected, the sign of the signal sharply changes when shifting the spot across the neighboring arms. 

The smooth relaxation function we observe in our universal electrical or optical quench-switching  is unparalleled in traditional magnetic devices in which the memory loss has a form of stochastic fluctuations between the two equilibrium states with opposite magnetization. Our quench-switching and relaxation functionality can be implemented  in analog spiking neuromorphic circuits  to mimic the neuron's leaky integration (Fig.~3b and Figs.~S3-7) and to detect the pulse order and delay (Figs.~3c,d and Figs.~S8,9) which determine the synaptic weights\cite{WulframGerstner2002,Kurenkov2019}. We have demonstrated that, at room temperature, our analog devices can simultaneously operate  in two relaxation ranges of milliseconds and seconds. The ranges can be further modified by, e.g.,  changing the operation temperature. We note that in the alternative magnetic memory concept considered for spiking neuromorphics prior to our work\cite{Kurenkov2019},  the time-dependence was realized via the extrinsic transient heating by the pulses. This is fundamentally distinct from our mechanism relying on the intrinsic smooth, universal and tunable relaxation from the higher energy metastable states. 

\bigskip
\noindent{\large\bf Quench-switching mechanism}

\noindent
In this section we discuss the microscopic origin of our new switching mechanism. In a parallel study on our CuMnAs devices\cite{Wornle2019}, electrical pulsing has been performed simultaneously with a scanning NV-diamond magnetometry  of antiferromagnetic domains  using a pump-probe scheme.  The imaging reveals a nano-scale fragmentation of the antiferromagnetic domains linked to the analog switching signals. The domain fragmentation, which is estimated to reach a 10~nm scale, is controlled by the number and amplitude of the pulses. Images  acquired at different pump-probe delays reveal that the fragmented domain pattern maintains a memory of the pristine state towards which it relaxes. These characteristics are mirrored in the multi-level switching and relaxation  of the electrical readout signal.  An independent confirmation of the pulse-induced fragmentation of antiferromagnetic domains by our XMLD-PEEM measurements is presented in Fig.~S14 (see also Methods).

Additionally,  a collection of structural characterization measurements has been performed on our samples, some presented in Supplementary Information Figs.~S13 and S14, and an  extensive set of results summarized in a parallel study\cite{Krizek2020}. The experiments show that the single-crystal CuMnAs epilayers grown with the best substrate matching, stoichiometry, uniformity, and minimized abundance of defects have optimized characteristics of our quench-switching signals. As mentioned above, we experimentally determined that the system heats up to the vicinity of the antiferromagnetic transition temperature during the short writing pulse. In comparison, the range of optimal growth temperatures (190-260$^\circ$C) exceeds the N\'eel temperature and at these high temperatures and for the much longer growth time (tens of minutes) than the pulse length, the synthesized epilayer preserves full crystal order. Our structural characterizations also include scanning electron microscopy (Figs.~S13) and X-ray absorption PEEM (Fig.~S14) on pulsed devices, transmission electron microscopy on thin lamellae while heated to temperatures exceeding the transient heating in devices during the pulse\cite{Krizek2020}, and X-ray diffraction after post-growth annealing of the crystal at these excessive temperatures (Fig.~S13). We observe no signs of electromigration or structural transition of the crystal for electric field and heating conditions corresponding to the pulsing experiments reported in our work.   

The transition to the metastable nano-fragmented domain state is, on the other hand, firmly established by the two independent magnetic microscopies. The excitation of the antiferromagnet into a fragmented domain state when bringing the system close to the N\'eel transition point can be readily ascribed to the entropy gain in the free energy \cite{Yin-Yuan1956}. Remarkably, our experiments show that the antiferromagnet retains the high resistive nano-fragmented domain state over time scales that  exceed by many orders of magnitude the time over which the system is in the transient excited state at elevated temperature. 

Regarding the magnitude of the  resistive signal we refer, for example, to an earlier study of the striped domain phase of a Co film \cite{Gregg1996} where a 5\% increase of the resistivity compared to the uniform magnetic state was caused by a density of domain walls about twenty-times lower than the density reached in the present experiments,  as inferred from the parallel NV-diamond magnetometry study on our devices \cite{Wornle2019}. Resistive changes of $\sim$100\% in our nano-fragmented state are, therefore, in line with the measurements in ferromagnets when extrapolated to our high domain wall densities. Another comparison is provided by earlier ab initio transport calculations in CuMnAs \cite{Maca2017}. Here a relative increase of the resistivity by over 100\% was obtained when comparing the resistivity of a material with a given substitutional lattice disorder and a perfect magnetic order to a material with the same lattice disorder and a strong frozen magnetic disorder.

Finally, we remark that in current-polarity dependent switching measurements, the NV-diamond magnetometry study also illustrates a coexistence of  the domain fragmentation with the N\'eel vector reorientation in the antiferromagnetic domains \cite{Wornle2019}. The N\'eel vector reorientation was also directly observed in earlier XMLD-PEEM studies\cite{Grzybowski2017,Wadley2018}. From the magnetic imaging and from another parallel study of AMR in CuMnAs in strong magnetic fields \cite{Wang2020} we conclude that for the resistive switching ratios on the order of a fraction of a per cent\cite{Wadley2016,Olejnik2018}, both AMR due the N\'eel vector reorientation in the domain and the resistivity increase due to the domain fragmentation can contribute.  With increasing amplitude of the resistive readout signal, the domain fragmentation quench-switching takes over. We note that the magnetic-field dependent AMR and XMLD \cite{Wang2020} indicate that the reorientation of the N\'eel vector inside the antiferromagnetic domains can occur at around 2~T while XMLD-PEEM measurements\cite{Wang2020} showed that the system remained in a multi-domain state even after applying and removing a 7~T field. Consistently with these microscopies, our large quench-switching signals, ascribed to nano-fragmented multi-domain states, are insensitive to strong magnetic fields (see Supplementary Information Fig.~S15 for our measurements in 14 T). 

\bigskip
\noindent{\large\bf Conclusions}

\noindent We have reported the quenching of an antiferromagnet into high-resistive nano-fragmented domain states after bringing the system close to the N\'eel transition point using electrical or ultra-short optical pulses. Key questions do though remain, which are related to the limiting excitation  time-scales of the quench-switching process, as well as the scattering mechanism responsible for the high resistivity of the quenched state. Furthermore, the processes that allow a single-crystal epilayer with a collinear antiferromagnetic ground state to remain in the fragmented metastable state for macroscopic timescales are currently unclear. To address these questions, considering the nanoscale fragmentation and the picosecond relaxation attempt times we observe, microscopic imaging techniques with higher spatial and temporal resolution will be needed. Furthermore, since electronic (structural) fluctuation at the nanoscale might accompany the strong magnetic disorder, both charge and spin sensitive imaging will be desirable to elucidate the microscopic origin of the activation energy barriers determining the retention/relaxation time-scale. The quench switching in dipolar-field free antiferromagnets reported here has no counterpart in ferromagnetic systems and could prove to be of value in a range of areas, from unconventional memory-logic devices to ultrafast optical switching. 

\protect\newpage

\bigskip
\noindent{\large\bf Methods}

\noindent
{\bf Growth.} of the films of tetragonal CuMnAs used in this study was done by molecular beam epitaxy at substrate temperatures around 200°C. The films were capped by 3nm of Al layer which, after quick oxidation following removal from vacuum, prevents oxidation of CuMnAs\cite{Krizek2020}. To pattern the devices we used electron beam lithography and wet chemical etching: Aluminium cap was removed in 2.7\% C$_4$H$_{13}$NO (3~s) and as the CuMnAs etchant we used a mixture 4:1:2 C$_4$H$_6$O$_6$(5\%):H$_2$O$_2$(5\%):H$_2$SO$_4$(10\%) with the rate of 50~nm per 30~s. 

\bigskip
\noindent
{\bf Synchronous electrical and optical readout.} (Figs.~4c,d) we performed using a fast NI data acquisition card which recorded simultaneously in one channel electrical resistance and in second channel analog output of a lock-in amplifier monitoring reflected light intensity (modulated by a photoelastic modulator at frequency 100~kHz).

\bigskip
\noindent
{\bf ns electrical pulses.}  (Fig4e, Fig.~S12) we delivered using a custom designed four-terminal high frequency PCB sample holder. The voltage pulses were generated by a specialized pulse generator (Kentech RTV-30 Sub-ns Pulser).

\bigskip
\noindent
{\bf 100~fs optical pulses.} (Figs.~5,6) were generated by a Ti:sapphire laser. The base repetition rate was 82~MHz and we utilized a pulse picker to obtain lower repetition rates. The wavelength of the laser light was 800~nm, corresponding to energy of 1.55~eV, i.e., well below the band gap of the GaP substrate (2.26~eV).

\bigskip
\noindent
{\bf Photo-emission electron microscopy (PEEM) images.} (Fig.~S14) were obtained on beamline I06 at Diamond Light Source, using an Elmitec SPELEEM-III microscope to image secondary-electron emission arising from X-rays incident on the sample at a grazing angle of 16$^\circ$. The probe depth was $\approx7$~nm, and the lateral resolution was $\approx50$~nm. A pulsing power supply was connected to the sample in ultrahigh vacuum via feedthroughs to the sample holder. Magnetic contrast images were obtained using the X-ray magnetic linear dichroism (XMLD) asymmetry, which measures the local surface N\'eel vector.  We acquired images for each X-ray energy during 1s exposure times, and we averaged 20 images to obtain a single XMLD-PEEM image. For more details see Ref.~\onlinecite{Wadley2018}.

\bigskip
\noindent
{\bf Data availability statement.} The data that support the plots within this paper and other findings of this study are available from the corresponding author upon reasonable request.

\bigskip
\noindent{\large\bf Acknowledgements}

\noindent
We acknowledge Ji\v{r}\'{\i} Kastil and Martin M\'{\i}\v{s}ek for experimental support. This work was supported in part by the Ministry of Education of the Czech Republic infrastructures Grants CzechNanoLab  No. LM2018110, No. LNSM-LNSpin and MGLM No. LM2018096, the Czech Science Foundation Grant No. 19-28375X,  the Charles University Grant GA UK No. 886317 and 1582417, the EU FET Open RIA Grant No. 766566, the Engineering and Physical Sciences Research Council Grant No. EP/P019749/1. P.W. acknowledges support from the Royal Society through a University Research Fellowship. T.J acknowledges the support from the Neuron Foundation Prize, and K.O. form the Neuron Foundation Impuls Grant.

\bigskip
\noindent{\large\bf Author Contributions}

\noindent
Author contributions: K.O., Z.K.,  and T.J. conceived and designed the project. Z.K., M.S., J.Z., F.K., K.O., P.W., K.W.E., S.R., O.J.A., F.M., and  S.S.D. performed experiments. K.O., P.N., P.W., K.W.E., J.W., X.M., M.S.W, P.G., and T.J. analyzed data. V.N., F.K., and R.P.C. contributed materials.  K.O. and T.J. wrote the paper.

\bigskip
\noindent{\large\bf Competing interests}

\noindent
The authors declare no competing interests.

\bigskip
\noindent{\large\bf Additional information}

\noindent
{\bf Supplementary Information} is available for this paper at https://doi.org/10.1038/s41928-020-00506-4.

\bigskip
\noindent
{\bf Reprints and permissions information} is available at www.nature.com/reprints. 

\bigskip
\noindent
{\bf Correspondence and requests for materials} should be addressed to K.O. and T.J.

\bigskip
\protect\newpage
\section*{Figures}

\begin{figure}[h!]
\begin{center}
\hspace*{-0cm}\epsfig{width=1\columnwidth,angle=0,file=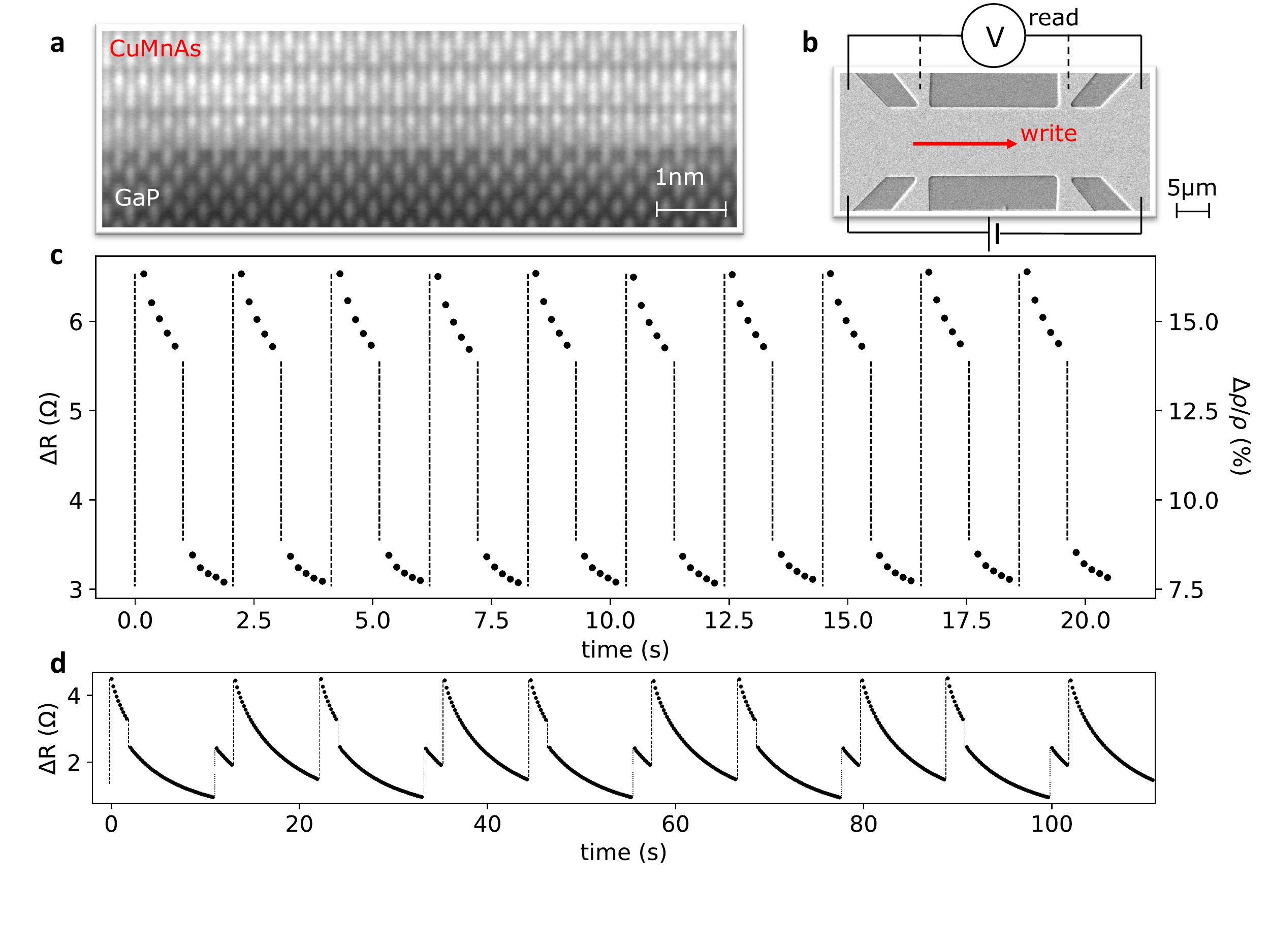}
\end{center}
\vspace*{-1cm}
\caption{{\bf Unipolar  high resistive switching in an elementary bar resistor microdevice.} {\bf a,}~TEM micrograph \cite{Krizek2020} of the single-layer CuMnAs antiferromagnet deposited on an insulating GaP substrate. {\bf b,}~SEM micrograph of the microbar device and schematics of the measurement set-up. Solid/dashed voltage probes correspond to the two/four-point readout measurement. {\bf c,}~Two-point measurement of alternating switching between higher and lower resistance metastable states at room temperature. The higher resistance state is written by a current amplitude of $1.2\times10^7$Acm$^{-2}$. The lower resistance state is written by a current pulse of a 8\% weaker amplitude. The electrical pulses have the same polarity and are 100~$\mu$s long. Dashed lines in {\bf c} are guides to the eye. {\bf d,}~A sequence of switching to high/low/low/high resistance states repeated five times. The higher resistance state is written by a current amplitude of $1.13\times 10^7$~Acm$^{-2}$ and the lower resistance state  by  $1.08\times 10^7$~Acm$^{-2}$.}
\label{fig1}
\end{figure}

\protect\newpage
\begin{figure}[h!]
\begin{center}
\hspace*{-0cm}\epsfig{width=1\columnwidth,angle=0,file=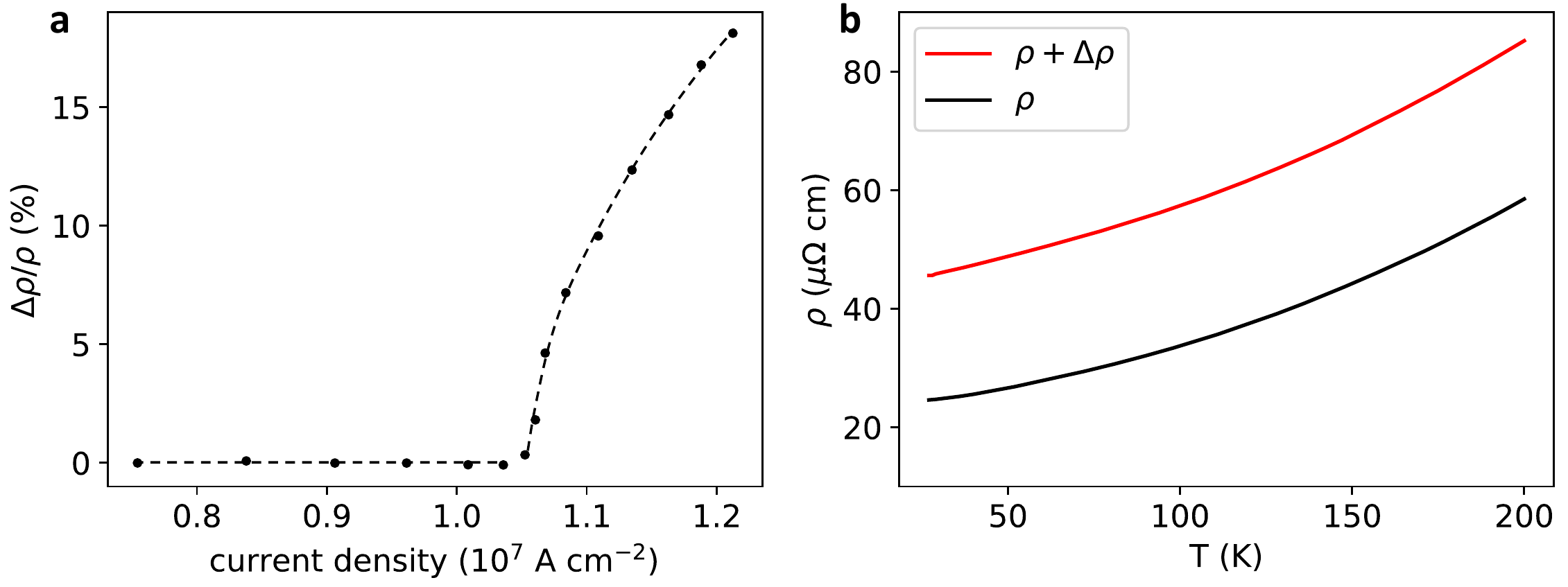}
\end{center}
\caption{{\bf Resistivity switching ratios.} {\bf a,}~Four-point measurement of the resistivity switching ratio as a function of the current density at room temperature. Dashed lines in {\bf a} are guides to the eye. {\bf b,}~Electrical switching is performed at 200~K and then the temperature dependence  of the resistivity after switching ($\rho + \Delta\rho$) is compared to the temperature dependence of the resistivity without applying the switching pulse ($\rho$).
}
\label{fig2}
\end{figure}

\protect\newpage
\begin{figure}[h!]
\begin{center}
\hspace*{-0cm}\epsfig{width=1\columnwidth,angle=0,file=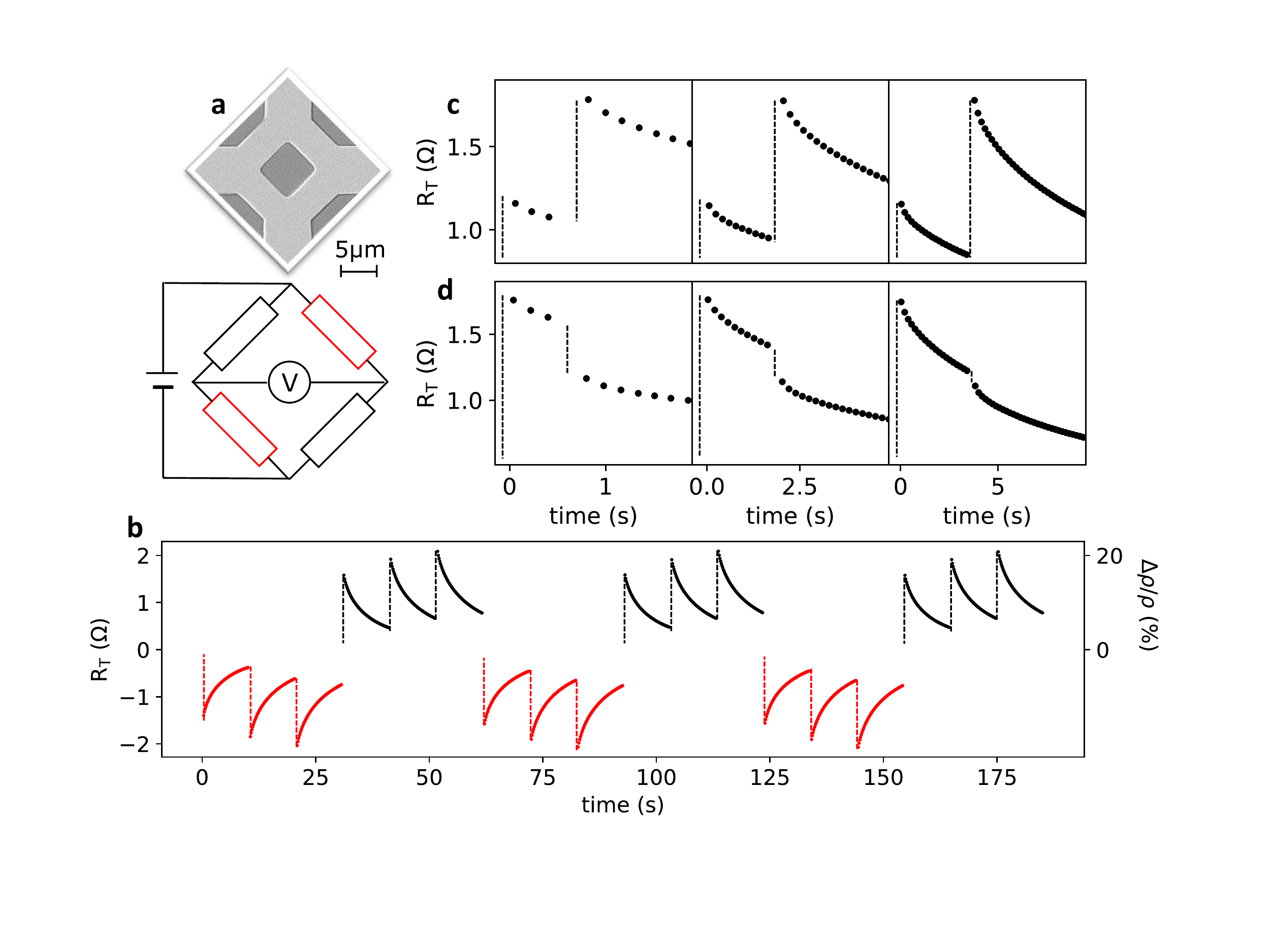}
\end{center}
\vspace*{-1cm}
\caption{{\bf Reproducible analog switching characteristics.} {\bf a,}~SEM micrograph of the Wheatstone-bridge device and schematics of the electrical writing applied along one pair of resistors (red) and the orthogonal pair of resistors (black). In the bridge readout set-up, the readout current is applied between two corners of the bridge (battery sign) and the voltage is measured between the other two corners (voltmeter sign). {\bf b,}~Switching signal measured across the bridge for three successive writing pulses of the same amplitude delivered along one pair of arms (red arrows in {\bf a,}) followed by three successive writing pulses of the same amplitude along the other pair of arms (black arrows in {\bf a,}). {\bf c,}~Lower/higher switching sequence for 9.5/10.5~V writing electrical pulses applied across the respective arms of the Wheatstone device (pulse-voltage includes the contact resistance contribution). The three panels correspond to three different delays between the lower and higher writing pulse. {\bf d,}~Same as {\bf c,} for the reversed order of the two pulses.
}
\label{fig3}
\end{figure}

\protect\newpage
\begin{figure}[h!]
\begin{center}
\hspace*{-0cm}\epsfig{width=1\columnwidth,angle=0,file=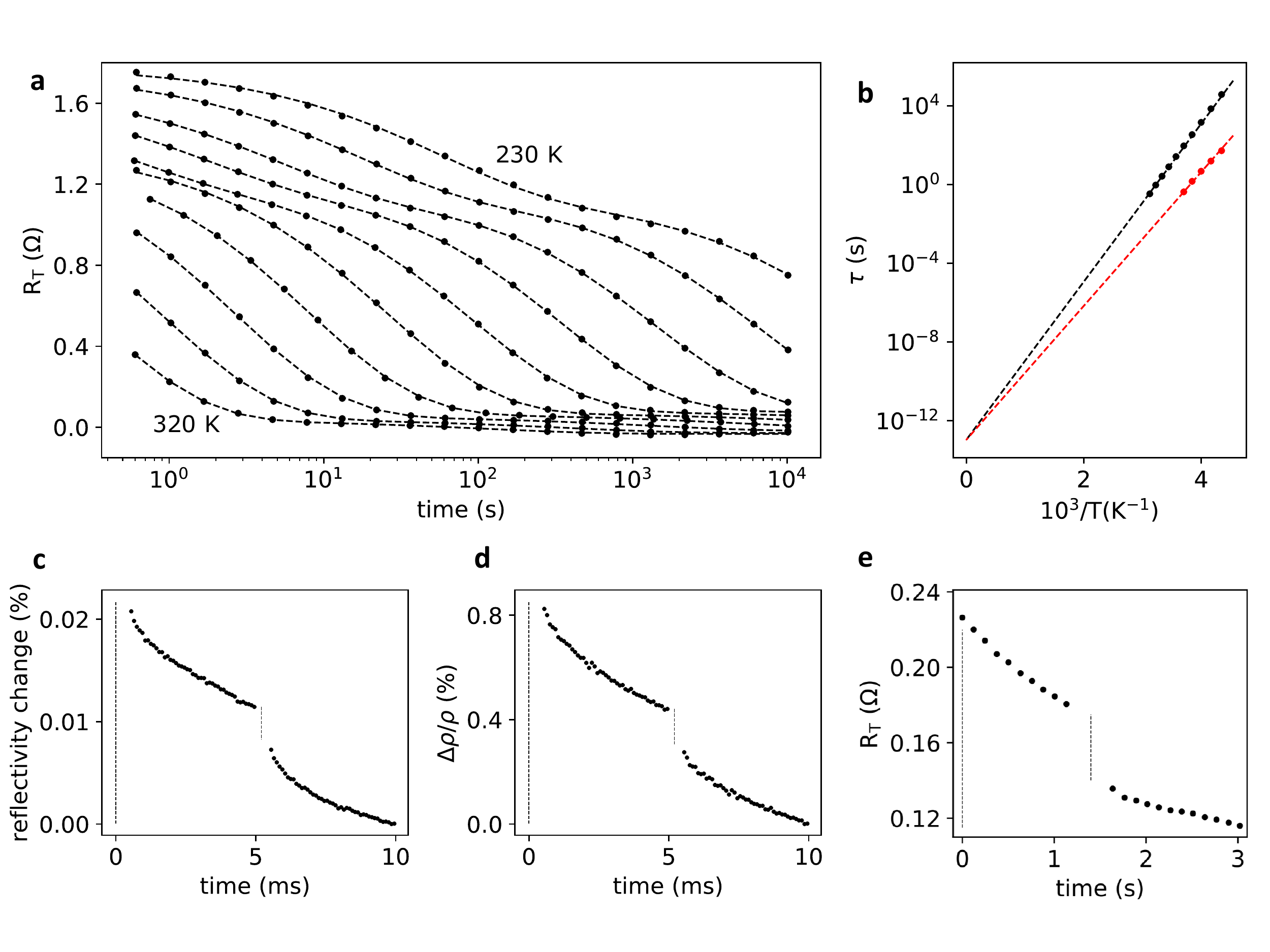}
\end{center}
\caption{{\bf Relaxation of the switching signal, optical readout, and switching by  ns pulses.} {\bf a,}~Time dependence of the relaxation of the switching signal in the Wheatstone-bridge device for temperatures ranging from 320 to 230~K (10~K step). Dashed lines are Kohlrausch stretched-exponential fits. {\bf b,}~Relaxation times $\tau_{1(2)}$ inferred from fits in {\bf a,} as a function of the inverse temperature. Dashed lines in {\bf a,b} are fits; the extrapolated $y$-axis intercept in {\bf b,} corresponds to the attempt time $\tau_0$ and the slopes to the activation barriers $E_{1(2)}$ (see text).  {\bf c,d}~Comparison of optical reflectivity and electrical resistivity measurements of the same higher/lower electrical switching sequence. The measured millisecond-range relaxation at room temperature corresponds to the faster relaxing component ($\tau_2$) from {\bf a,}. {\bf e,}~Room-temperature electrical detection of a higher/lower electrical switching sequence for 1~ns long pulses of amplitude 1.3  and  $1.1\times10^8$Acm$^{-2}$, resp.}
\label{fig4}
\end{figure}

\protect\newpage
\begin{figure}[h!]
\begin{center}
\hspace*{-0cm}\epsfig{width=1\columnwidth,angle=0,file=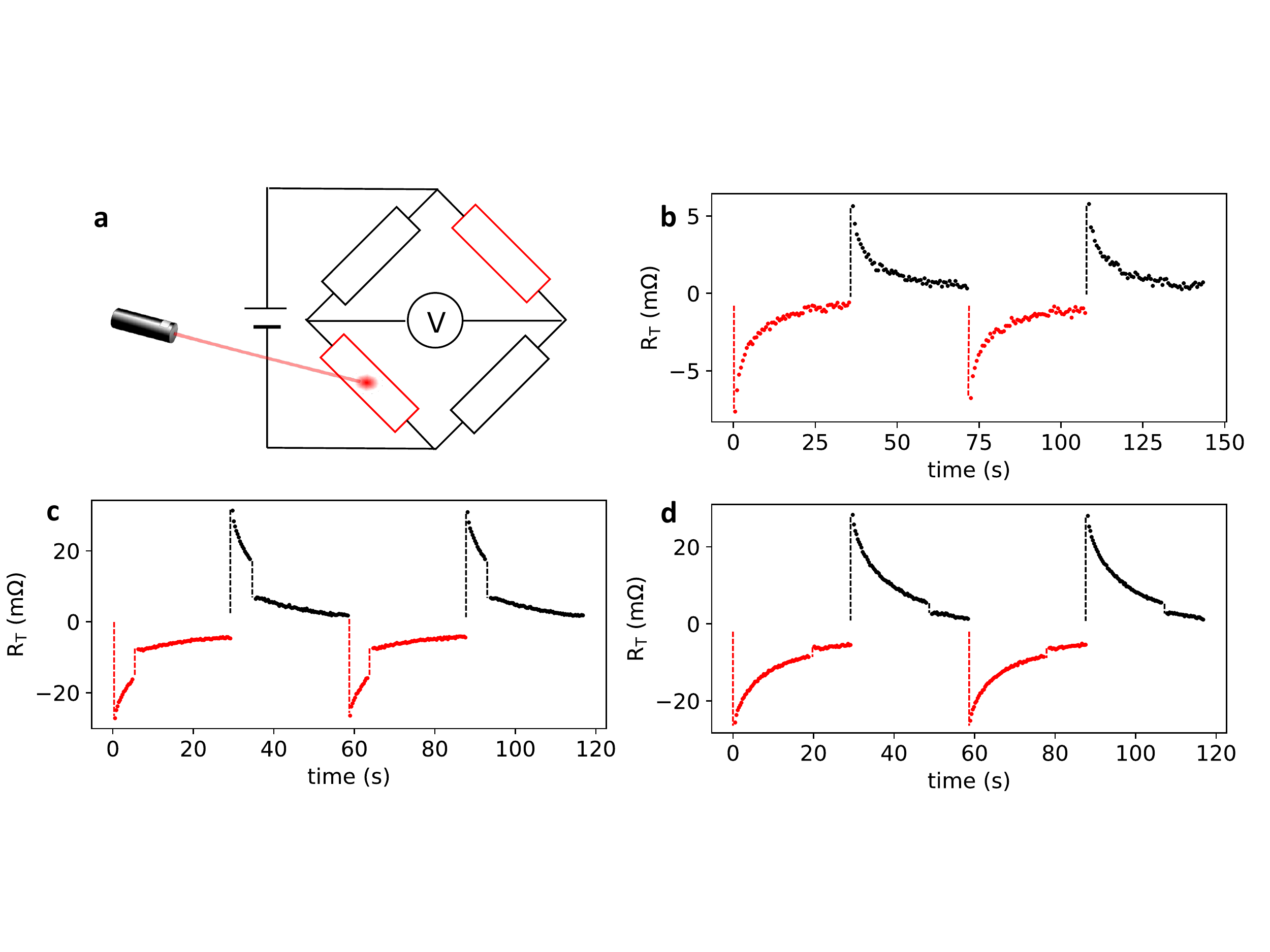}
\end{center}
\vspace*{-1cm}
\caption{{\bf fs-laser pulse switching.} {\bf a,} Schematics of the optical writing by a laser spot focused on one arm of the Wheatstone bridge (red resistor) and then on the orthogonal arm (black resistor). Electrical readout is performed as in Fig.~\ref{fig3}a.  {\bf b,}~Electrical readout signal recorded $550$~ms after optical switching by a single 100~fs, 1.7~$\mu$m-focus pulse  of 1~nJ delivered to one spot on one arm (red) and one spot on the neighboring arm (black) in a bridge device with 2.5~$\mu$m wide arms.  {\bf c,}~Optical writing by a laser beam of 10~kHz repetition rate over 250~ms and with 0.8~nJ per 100~fs pulse, focused to a single  1.7~$\mu$m-diameter spot in one arm (red). After 5~s, a second laser pulse was applied defocused  to a $2\times$ larger spot. The same procedure was then applied to the neighboring arm (black) in a bridge device with 5~$\mu$m wide arms. {\bf d,}~Same as {\bf c,} with the delay of 20~s between the first and second pulse. All measurements are at room temperature.
}
\label{fig5}
\end{figure}

\protect\newpage
\begin{figure}[h!]
\begin{center}
\hspace*{-0cm}\epsfig{width=1\columnwidth,angle=0,file=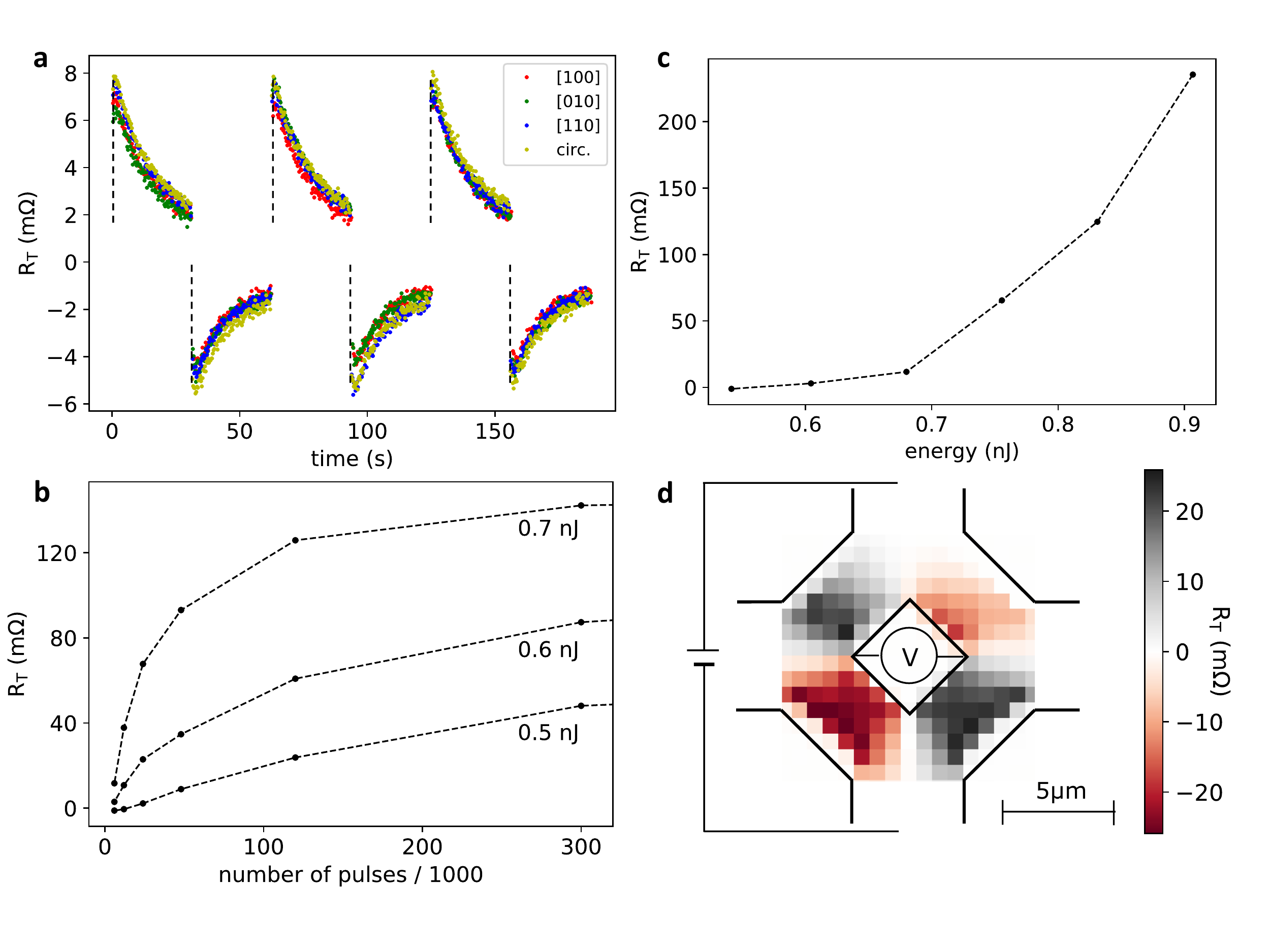}
\end{center}
\vspace*{-1cm}
\caption{{\bf Optical switching characteristics.} {\bf a,} Switching signals measured for linearly polarized laser pulses along specified crystallographic directions in the sample, and for circularly polarized laser pulses (circ.). Experiments were performed with 0.5~nJ per pulse, 10~ms pulse train, and 80~MHz repetition rate. {\bf b,}~Electrical readout signal as a function of the number of 100~fs laser pulses (with energy per pulse from 0.5 to 0.7~nJ) recorded $550$~ms after completing a 600~ms scan of the switching laser beam over one arm of the bridge with the repetition rate varied from 10~kHz to 0.5~MHz. {\bf c,}~Dependence of the optical switching signal on the energy per 100~fs laser pulse for a 600~ms long pulse train with 10~kHz repetition rate. {\bf d,}~Spatially resolved, pixel-by-pixel optical switching combined with electrical readout in the Wheatstone-bridge device. Each micron-size pixel was exposed to a train of pulses with a 10~kHz repetition rate over 250~ms. The beam was focused to a $\sim1$~$\mu$m spot and delivered energy of 0.7~nJ per 100~fs pulse. 
}
\label{fig6}
\end{figure}
%


\renewcommand{\figurename}{Figure S}
\setcounter{figure}{0}

\protect\newpage

\bigskip

\noindent{\large\bf Supplementary Note 1}

\pagenumbering{roman}
\setcounter{page}{1}

\noindent{\bf Electrical switching}

This section contains supplementary figures on analog switching characteristics for 100~$\mu$s electrical pulses (Figs.~S1, S3-S10), temperature rise during the pulse and current-voltage characteristic (Fig.~S2), fitting of the time and temperature dependent relaxation (Fig.~S11), and switching by 1~ns electrical pulses with comparison of switching energies for ns-electrical and fs-laser pulses (Fig.~S12).

\begin{figure}[h!]
\begin{center}
\hspace*{-0cm}\epsfig{width=1\columnwidth,angle=0,file=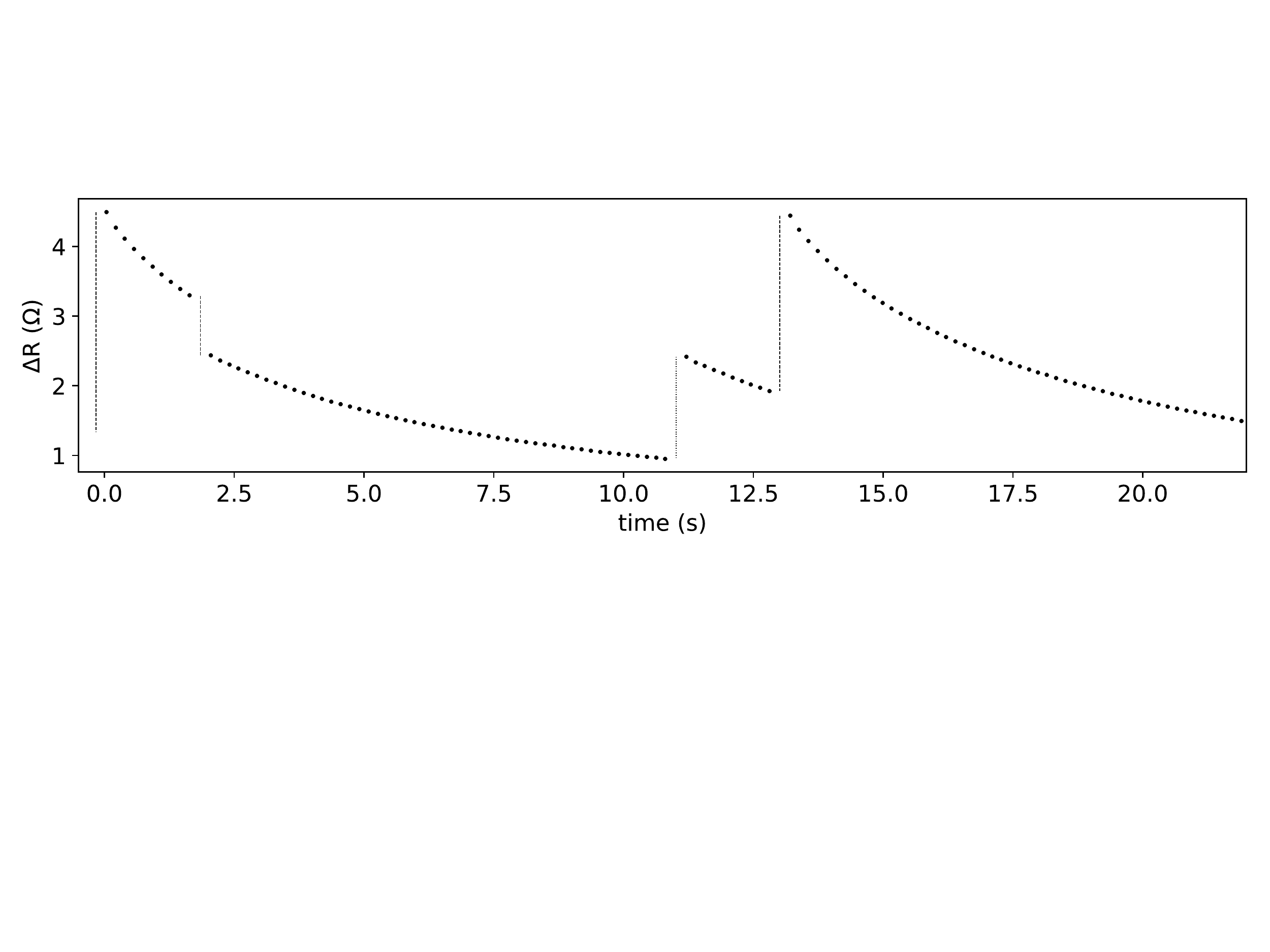}
\end{center}
\caption{A sequence of switching to high/low/low/high resistance states at room temperature in the microbar device (Fig.~1b in the main text). The higher resistance state is written by a current amplitude of $1.13\times 10^7$~Acm$^{-2}$ (corresponding to 7.1~V voltage drop between the 20~$\mu$m spaced voltage probes). The lower resistance state is written by a $1.08\times 10^7$~Acm$^{-2}$ current pulse (6.4~V). The current pulses have the same polarity and are 100~$\mu$s long. 
}
\label{figS1}
\end{figure}

\protect\newpage

\begin{figure}[h!]
\begin{center}
\hspace*{-0cm}\epsfig{width=1\columnwidth,angle=0,file=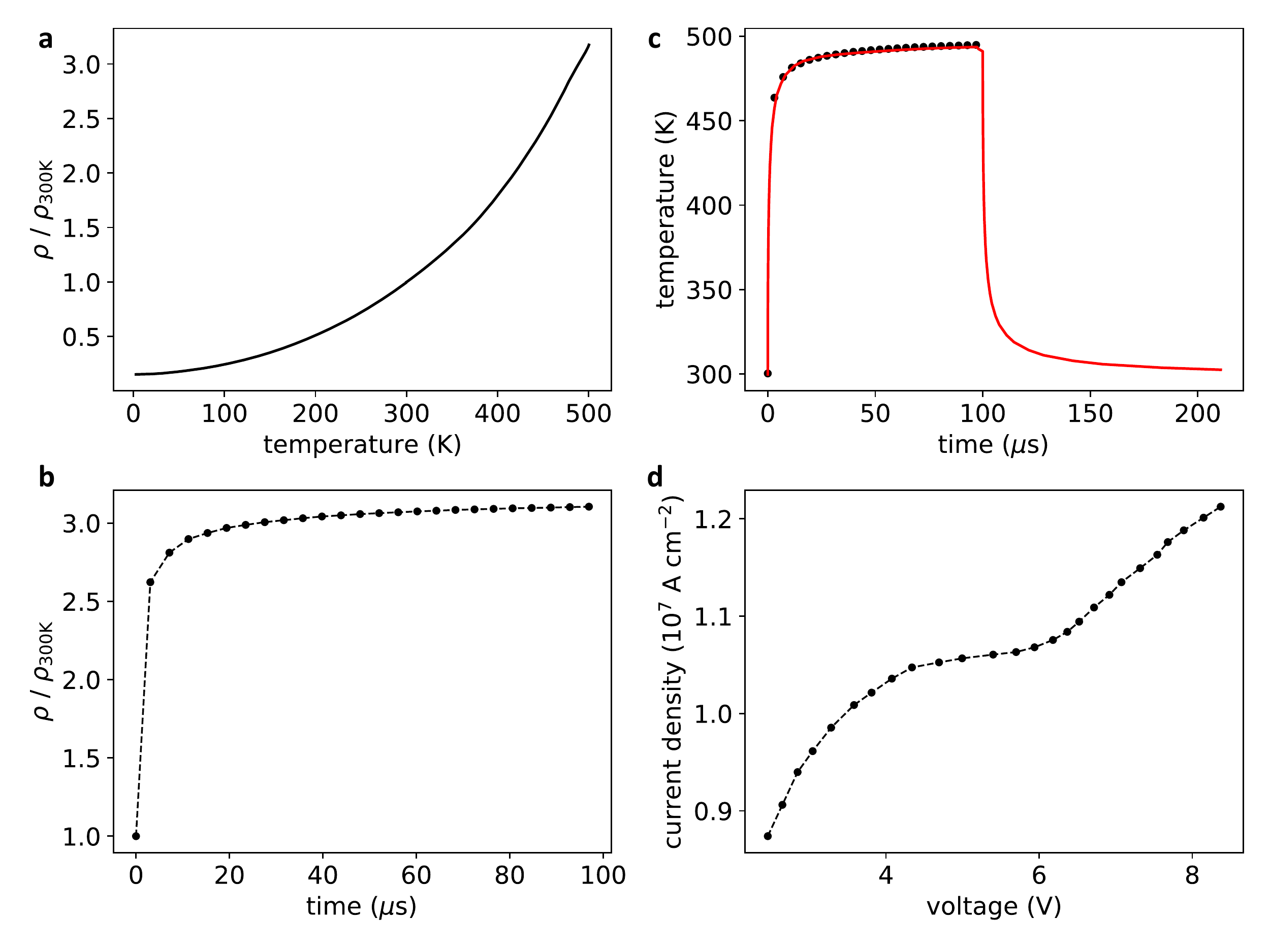}
\end{center}
\caption{{\bf a,} Measured temperature dependence of the CuMnAs resistivity normalized to the room temperature value. {\bf b,} Measured evolution of the resistivity  of the central part of the CuMnAs  microbar device  during the 100 $\mu$s writing pulse (normalized to the initial room temperature value before the start of the pulse). {\bf c,} Measured data obtained by combining panels {\bf a,b} and Comsol simulation of the time dependence of the  temperature rise during the 100~$\mu$s writing pulse and Comsol simulation of the decay after the pulse. {\bf d,} Pulsing current-voltage characteristics of the microbar device with the voltage measured between the 20~$\mu$m spaced voltage probes (see Fig.~1b in the main text).
}
\label{figS2}
\end{figure}

\protect\newpage

\begin{figure}[h!]
\begin{center}
\hspace*{-0cm}\epsfig{width=1\columnwidth,angle=0,file=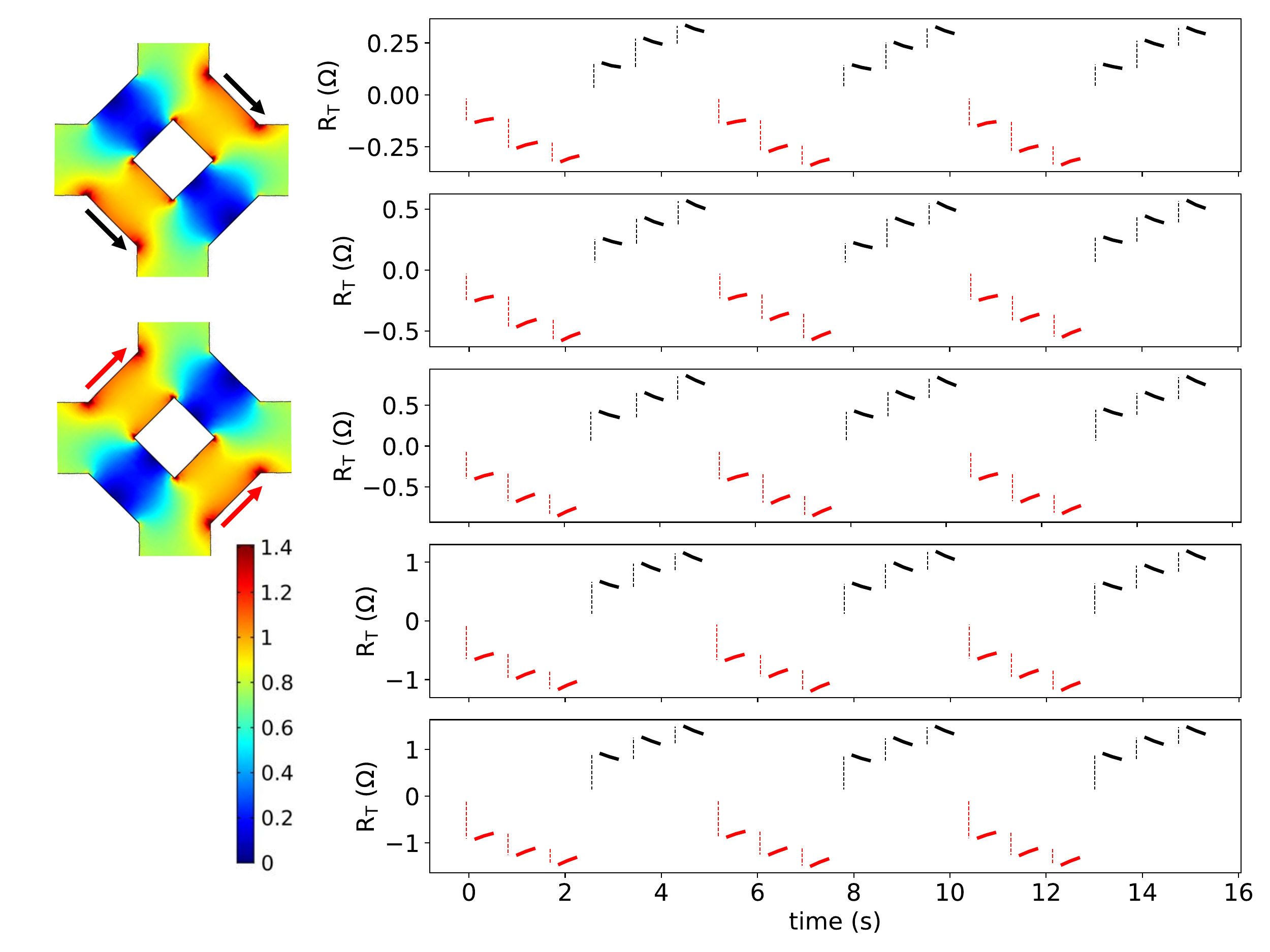}
\end{center}
\caption{Left: Comsol simulation of the writing current density map in the  Wheatstone bridge device.  Right: Switching signal measured across the bridge for three successive writing pulses of the same amplitude delivered along one pair of arms (red arrows in the left panel) followed by three successive writing pulses of the same amplitude along the other pair of arms (black arrows in the left panel). From top to bottom the voltage applied during the pulse across the respective arms (including the contact resistance contribution) is 9--11~V. The delay between successive pulses is 0.8~s. The experiments were done at room temperature.
}
\label{figS3}
\end{figure}

\protect\newpage

\begin{figure}[h!]
\begin{center}
\hspace*{-0cm}\epsfig{width=1\columnwidth,angle=0,file=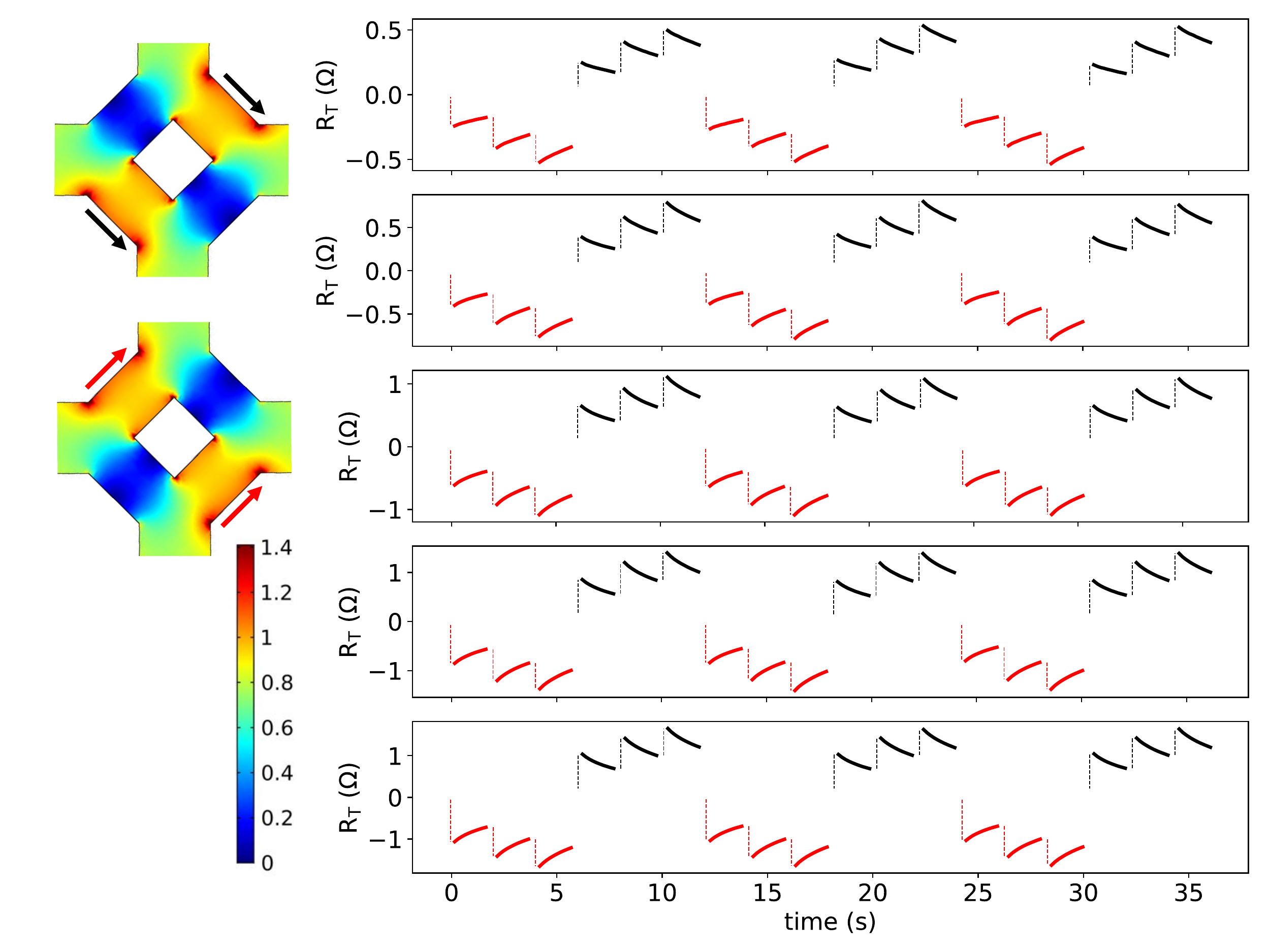}
\end{center}
\caption{Same as Fig.~S3 for the delay between pulses of 2~s.
}
\label{figS4}
\end{figure}

\protect\newpage

\begin{figure}[h!]
\begin{center}
\hspace*{-0cm}\epsfig{width=1\columnwidth,angle=0,file=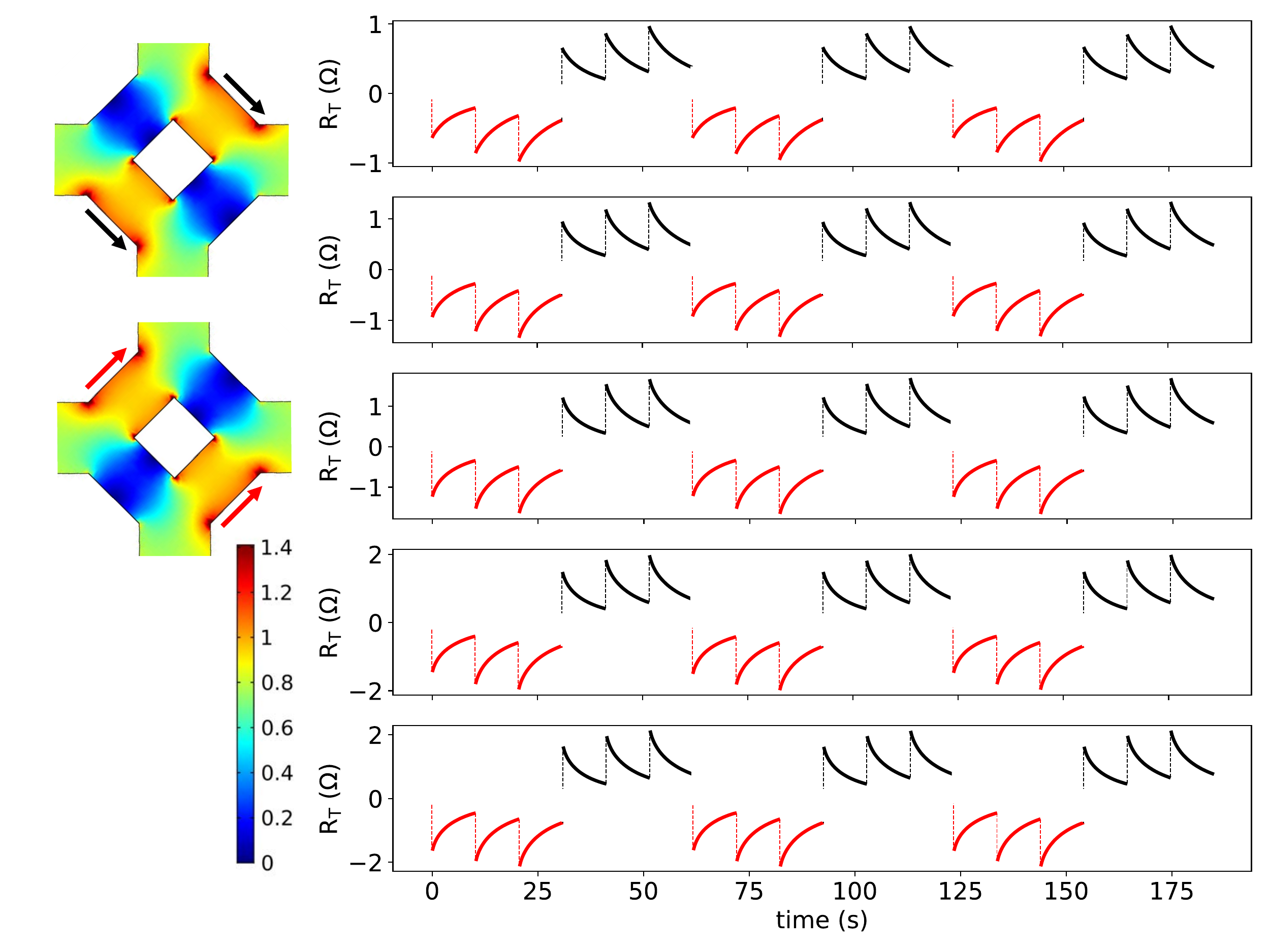}
\end{center}
\caption{Same as Fig.~S3 for the delay between pulses of 10~s.
}
\label{figS5}
\end{figure}

\protect\newpage

\begin{figure}[h!]
\begin{center}
\hspace*{-0cm}\epsfig{width=1\columnwidth,angle=0,file=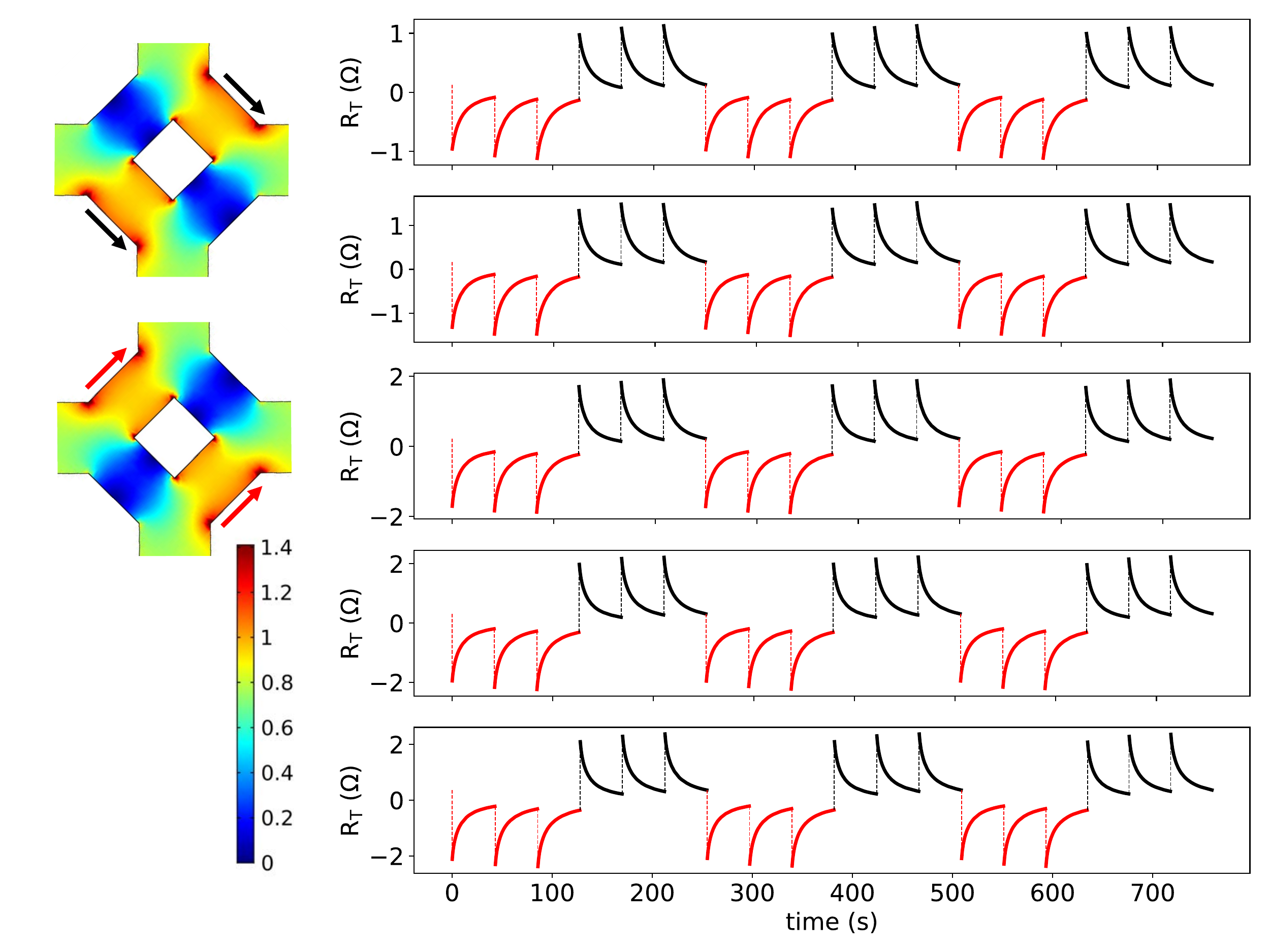}
\end{center}
\caption{Same as Fig.~S3 for the delay between pulses of 41~s.
}
\label{figS6}
\end{figure}

\protect\newpage

\begin{figure}[h!]
\begin{center}
\hspace*{-0cm}\epsfig{width=1\columnwidth,angle=0,file=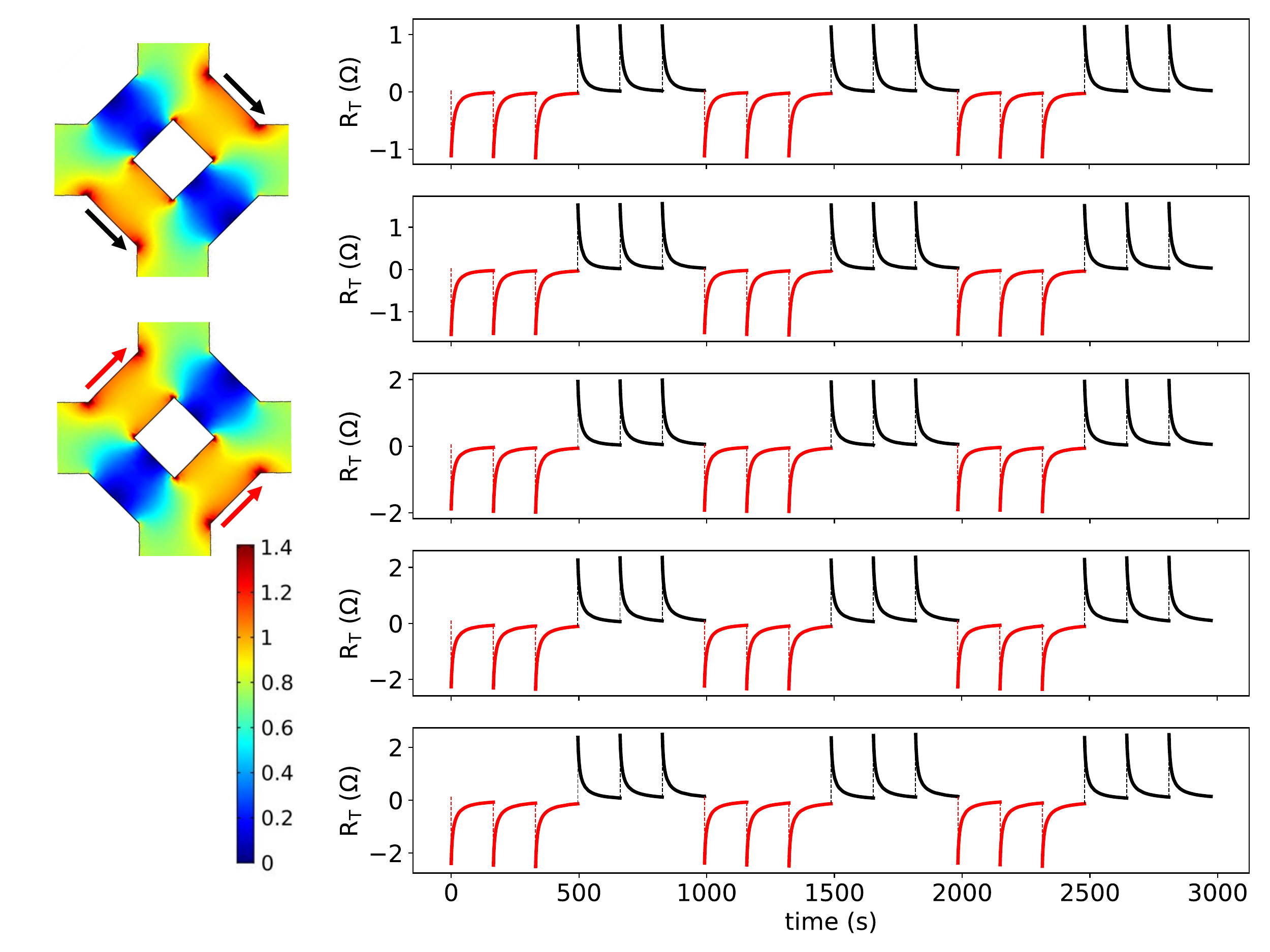}
\end{center}
\caption{Same as Fig.~S3 for the delay between pulses of 163~s.
}
\label{figS7}
\end{figure}

\protect\newpage

\begin{figure}[h!]
\begin{center}
\hspace*{-0cm}\epsfig{width=1\columnwidth,angle=0,file=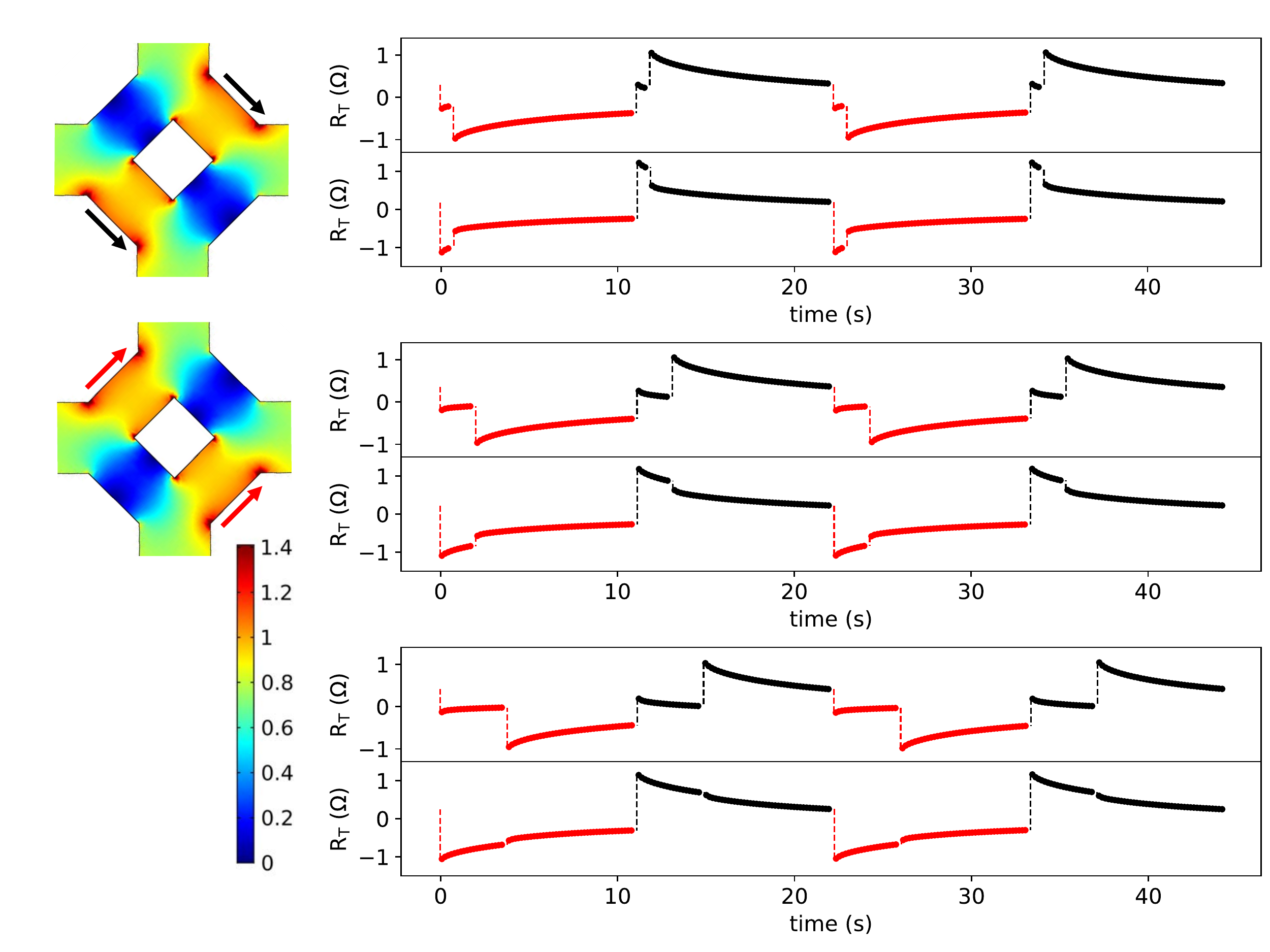}
\end{center}
\caption{Each double-panel: A sequence of lower/higher (top panel) and higher/lower (bottom panel) pulses applied first along one pair of arms of the Wheatstone bridge (red arrows) and then along the other pair (black arrows). The measurement is repeated twice in each double-panel and the double-panels differ in the delay between the two pulses in the sequence. The experiments were done at room temperature.
}
\label{figS8}
\end{figure}

\protect\newpage

\begin{figure}[h!]
\begin{center}
\hspace*{-0cm}\epsfig{width=.8\columnwidth,angle=0,file=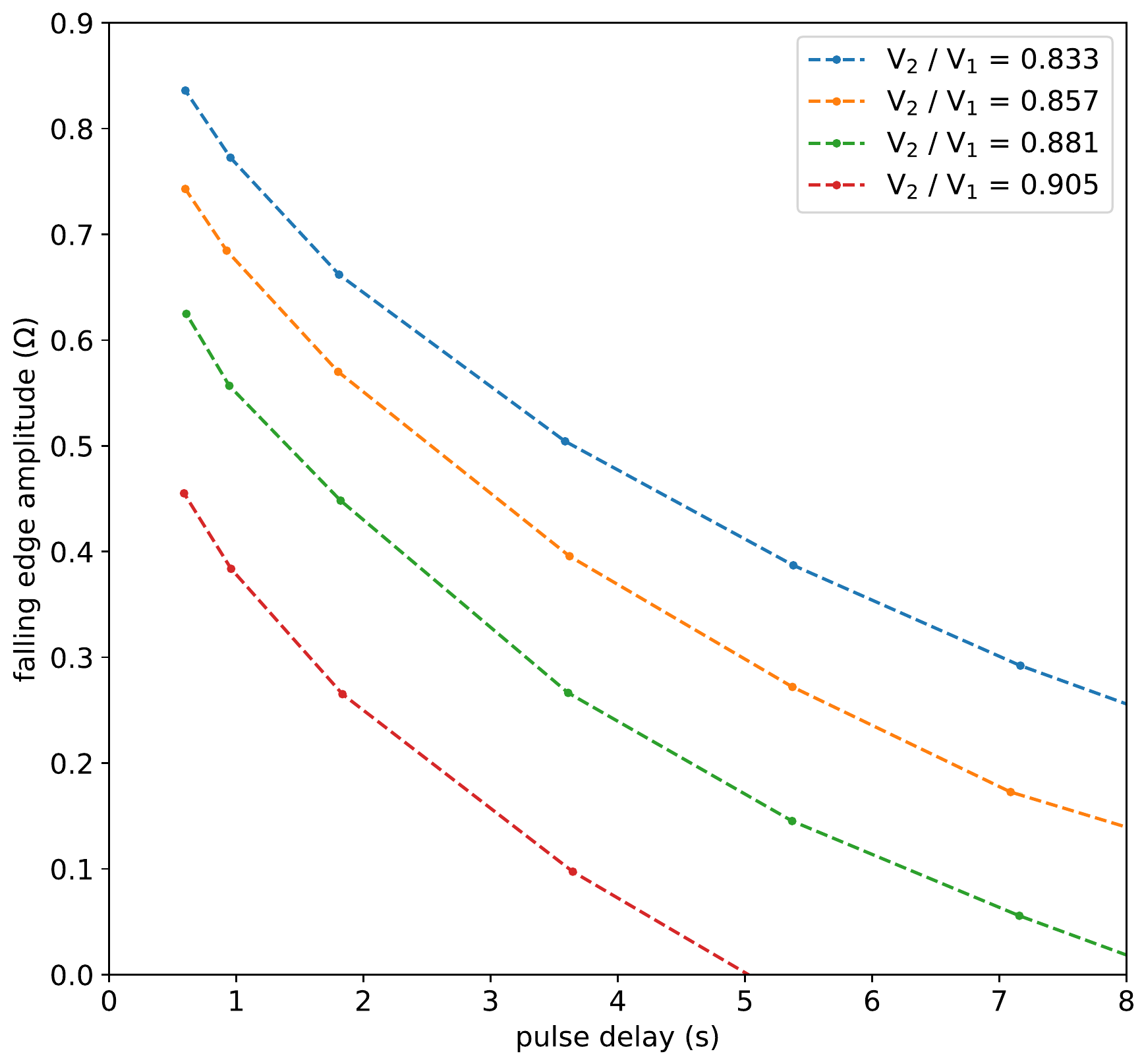}
\end{center}
\caption{The dependence of the amplitude of the falling edge for the higher/lower sequence of pulses as a function of the pulse delay for different ratios of the writing voltage pulses applied in the Wheatstone device at room temperature. 
}
\label{figS9}
\end{figure}

\protect\newpage

\begin{figure}[h!]
\begin{center}
\hspace*{-0cm}\epsfig{width=.8\columnwidth,angle=0,file=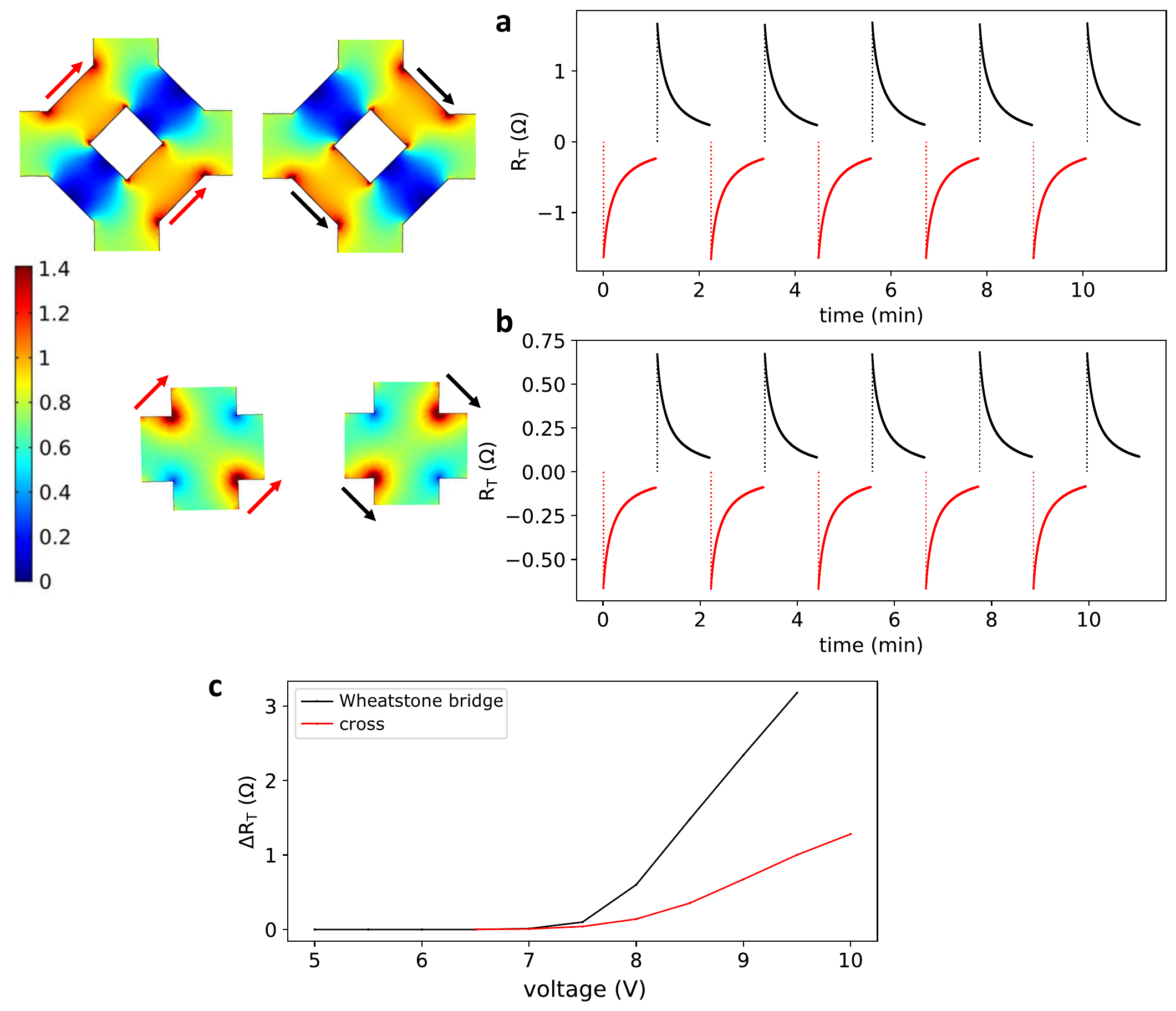}
\end{center}
\caption{{\bf a,} An example of the switching signal $R_{\rm T}$ measured on the Wheatstone bridge device  for 100~$\mu$s, 9.5~V  pulses  applied along one pair of arms (red arrows) and the orthogonal pair (black arrows). {\bf b,} Analogous signal measured on a cross device for or 100~$\mu$s, 10~V pulses. Left panels: Comsol simulations of the writing current density maps. {\bf c,} The amplitude of the switching signal as a function of the applied pulse voltage. In the Wheatstone bridge geometry, the amplitude of the signal can be directly related to the homogeneous change of the resistivity in the pulsed arms. In the  cross geometry, the resistance change occurs primarily in the corners exposed to higher current density than the interior of the cross. As a result, the value of the signal observed in the  cross geometry is significantly reduced because the readout electrical signal is effectively shunted by the central part of the device where switching is less efficient.  From the experiment we estimate that the maximum achievable signal in simple cross geometry is approximately 40\% of the value observed in the Wheatstone bridge for devices of size $\sim10$~$\mu$m.
}
\label{figS10}
\end{figure}

\protect\newpage

\begin{figure}[h!]
\begin{center}
\hspace*{-0cm}\epsfig{width=1\columnwidth,angle=0,file=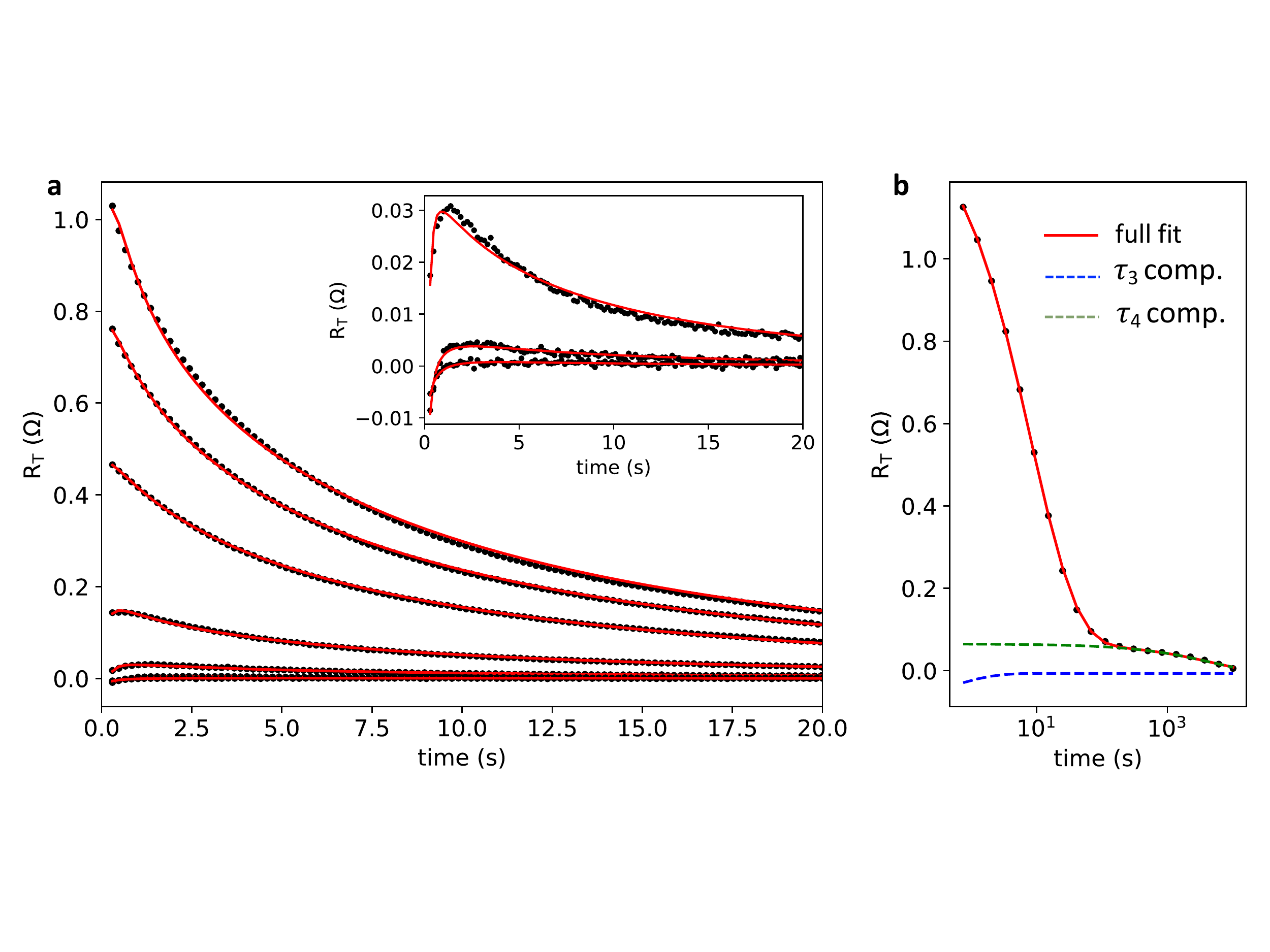}
\end{center}
\caption{{\bf a,} Stretched exponential fits and measured switching signals in the Wheatstone device at 290~K. From bottom to top, the curves correspond to pulse voltages 7.5--10.5~V. Inset: zoom-in for the voltages 7.5, 8, and 8.5~V highlighting the presence of an additional stretched-exponential component with a relaxation time $\tau_3=0.3$~s at 290~K and an opposite sign to the two main components discussed in Fig.~3 of the main text. {\bf b,} Data for the pulse voltage of 10.5~V for an enlarged and logarithmic time-scale highlighting the fourth stretched exponential component with $\tau_4=620$~s at 290~K. The  $\tau_{3,4}$ components are much weaker than the leading components with $\tau_{1,2}$ discussed in the main text.
}
\label{figS11}
\end{figure}

\protect\newpage

\begin{figure}[h!]
\begin{center}
\hspace*{-0cm}\epsfig{width=.5\columnwidth,angle=0,file=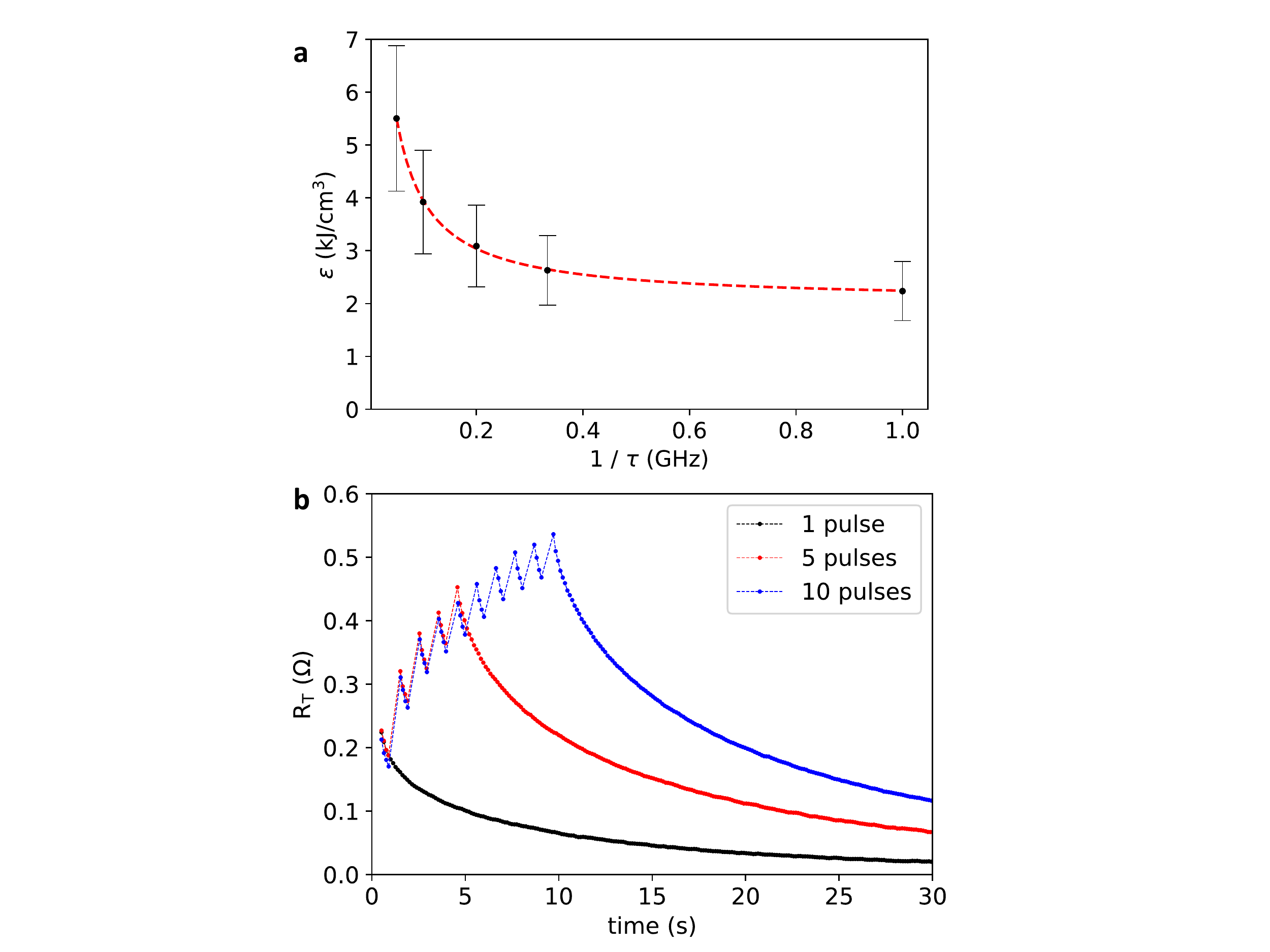}
\end{center}
\caption{{\bf a,}  Joule heating energy density as a function of the switching speed $1/\tau$ ($\tau$ is the pulse length) measured on a 5~$\mu$m wide bar device. The energy density is calculated from the current density ($0.9\times10^8$~Acm$^{-2}$ for $\tau=1$~ns) and resistivity during the pulse. At 200$^\circ$C, the resistivity increases by a factor of 3 compared to room temperature. The error bars correspond to the estimated temperature increase during the pulse of $200\pm50\circ$C. Measurements are performed close to the switching threshold with the corresponding switching signal of 1\%. For comparison, switching by the single 100~fs laser pulse of energy 1~nJ with the measured 30\% absorption in the sample, 1.7~$\mu$m spot size, and 50~nm film thickness (Fig.~4a of the main text) corresponds to energy density of 2.6~kJcm$^{-3}$. This is comparable to the saturated energy value for ns-electrical pulses. {\bf b,} Switching signal  after a single 1~ns pulse (black), and five (red) and ten (blue) 1~ns pulses of amplitude  $1.2\times10^8$Acm$^{-2}$. The experiments were done at room temperature.
}
\label{figS12}
\end{figure}

\protect\newpage

\noindent{\large\bf Supplementary Note 2}

\noindent{\bf Structural characterization and magnetic imaging}

This section contains figures with scanning electron micrographs of devices before and after pulsing and  X-ray diffraction of films before and after annealing (Fig.~S13), and X-ray absorption PEEM and XMLD-PEEM images of devices before and after pulsing (Fig.~S14). Below we also discuss parasitic effects reported elsewhere in sputtered Pt/insulator bilayers.


\begin{figure}[h!]
\begin{center}
\hspace*{-0cm}\epsfig{width=1\columnwidth,angle=0,file=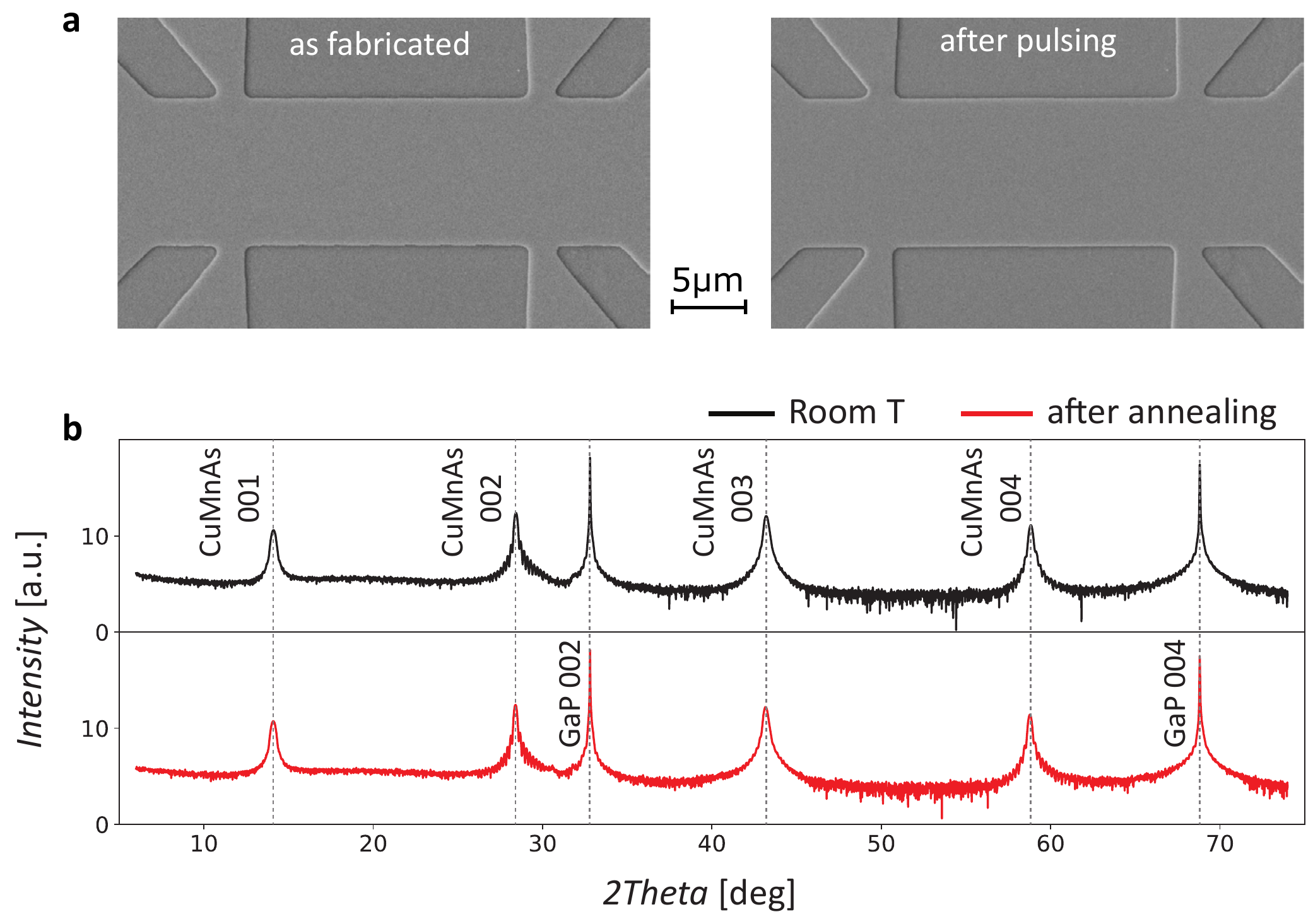}
\end{center}
\caption{{\bf a,} SEM micrograph of the  device as fabricated before electrical pulsing (left) and after pulsing (right) into a 16\% resistive switching signal at room temperature. {\bf b,} XRD $2\theta$ scan  of a 50 nm thick CuMnAs film grown on GaP measured at room temperature is shown in the top panel (black curve). The bottom panel shows a measurement by the same method for the same sample after three cycles of annealing in vacuum at 260$^\circ$C for 15 min and cooling back down 
to room temperature (red curve).  
}
\label{figS13}
\end{figure}

\protect\newpage

\begin{figure}[h!]
\begin{center}
\hspace*{-0cm}\epsfig{width=.8\columnwidth,angle=0,file=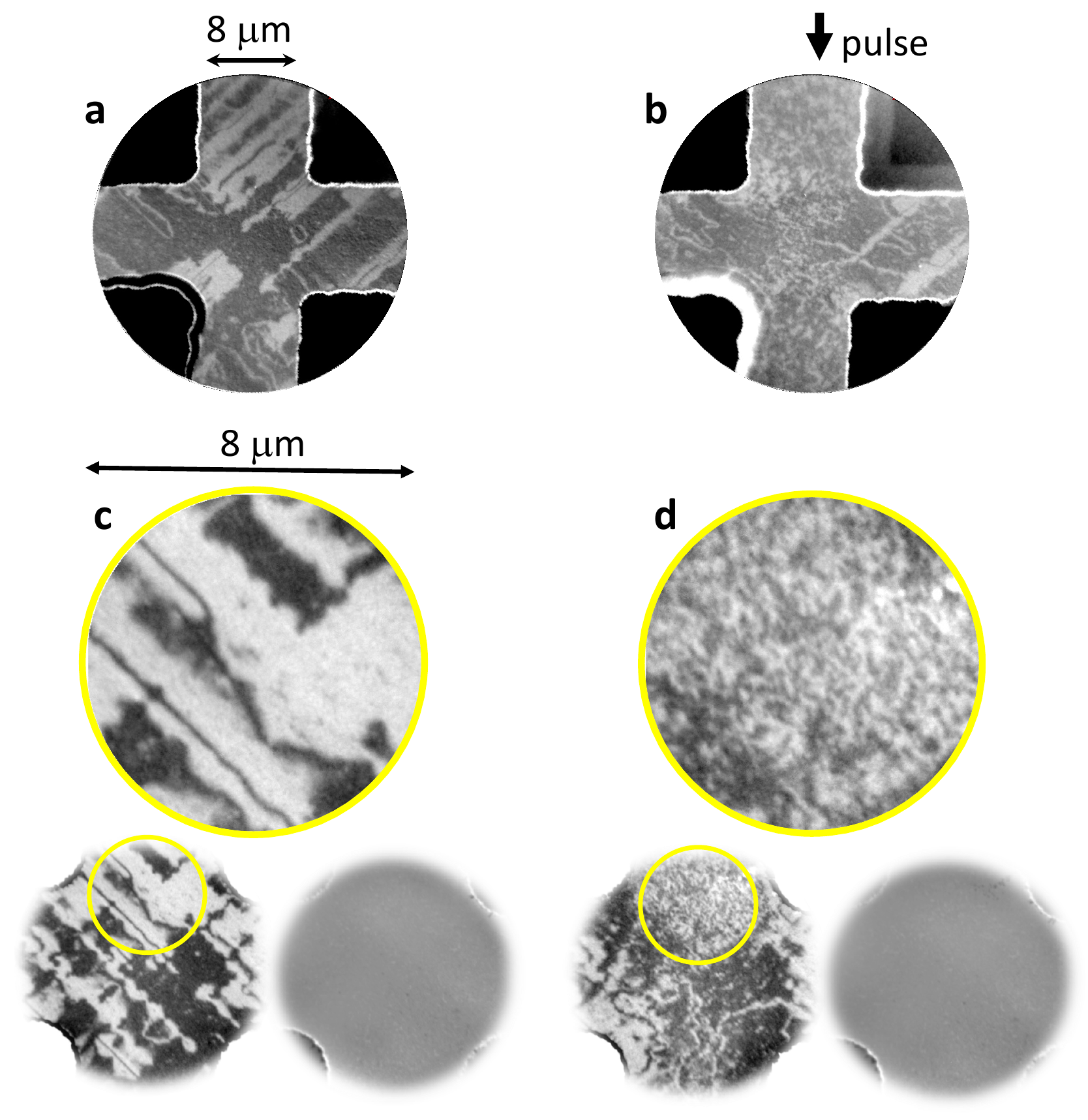}
\end{center}
\caption{\textbf{a,} Antiferromagnetic domain structure of an as-fabricated CuMnAs(50nm)/GaP sample observed by XMLD-PEEM for X-ray polarization $E \parallel [1\bar{1}0]$ crystal axis. The light (dark) contrast corresponds to antiferromagnetic domains with the N\'eel vector oriented parallel (perpendicular) to the X-ray polarization. \textbf{b,} Same as \textbf{a}, after applying a 500 $\mu$s current pulse of amplitude $\approx 1\times10^7$~Acm$^{-2}$ along the vertical ([010]-oriented) direction (black arrow). \textbf{c,} Bottom-left and top zoomed-in: XMLD-PEEM image of the antiferromagnetic domain structure of an as-fabricated CuMnAs(45nm)/GaP sample. Bottom right: X-ray absorption PEEM without XMLD contrast. \textbf{d,} Same as \textbf{c}, after applying a 100 ms current pulse of amplitude $\approx 1\times10^7$~Acm$^{-2}$ along the vertical ([010]-oriented) direction. For more details on the experiment see Methods in the main text and Ref.~\onlinecite{Wadley2018}.}
\label{figS14}
\end{figure}

\protect\newpage

Several papers have recently reported parasitic pulse-induced resistive changes in Pt-(antiferromagnetic)insulator bilayers with few-nm thick Pt films\cite{Chiang2019,Zink2019,Zhang2019c,Baldrati2019,Cheng2020,Churikova2020} These parasitic signals, typically of a fraction of per cent amplitude, are linked to the well-established elecromigration effects promoted at grain boundaries in microdevices of sputtered thin metal films\cite{Kozlova2013}. A Viewpoint article\cite{Zink2019} on Ref.~\onlinecite{Chiang2019} emphasizes that the possible occurrence of parasitic  effects in these samples highlights the importance of direct  imaging of the magnetic switching in the insulating antiferromagnet.

In contrast to the above studies, we use a single-crystal metallic antiferromagnet CuMnAs grown by molecular beam epitaxy. 
The switching to the metastable nano-fragmented antiferromagnetic domain state in our devices was verified by a parallel systematic laboratory study employing the NV-diamond microscopy \cite{Wornle2019}, and by our independent synchrotron XMLD-PEEM measurements (Supplementary Fig. S14).  Previous XMLD-PEEM measurements in CuMnAs also provided direct images of the earlier mechanism of the N\'eel vector reorientation controlled by the current direction\cite{Wadley2018}.

The absence of detectable structural transitions is discussed in the main text, evidenced here in Supplementary Fig.~S13 and S14, and in an extensive parallel study\cite{Krizek2020}. The magnetic transition to the metastable nano-fragmented domain state is, on the other hand, firmly established by the magnetic microscopies. Regarding the magnitude of the corresponding resistive signal we refer in the main text to the consistency with earlier domain wall resistance studies in ferromagnets\cite{Gregg1996} and ab initio transport calculations in CuMnAs\cite{Maca2017}. We also recall from the main text that we performed, in the same material and device structure, electrical and optical switching experiments with pulses covering the nine orders of magnitude range from microseconds to femtoseconds and that we observed precisely reproducible relaxation of the metastable state with attempt times in the picosecond range, consistent with the internal dynamics scale of antiferromagnets. 

Unlike the claims in Ref.~\onlinecite{Chiang2019} on their experiments on sputtered polycrystalline Pt-insulator structures, we experimentally excluded the possibility that our signals are due transient heating induced by the writing pulse and persisting during readout. First evidence is the resistance drop after the weaker pulse (Figs.~1-4 in the main text and Supplementary Figs.~S1 and S8), i.e., that we can deterministically control the switching towards not only the high but also the low resistivity states by the pulse amplitude. Second, we showed in Supplementary Fig. S2 that within a few $\mu$s after the $100\mu$s pulse, the sample cools down back to the base temperature. Our switching  signals are therefore detected at times safely exceeding the transient heating time.

\bigskip \bigskip

\noindent{\large\bf Supplementary Note 3}

\noindent{\bf Switching in the presence of strong magnetic fields}

\begin{figure}[h!]
\begin{center}
\hspace*{-0cm}\epsfig{width=.6\columnwidth,angle=0,file=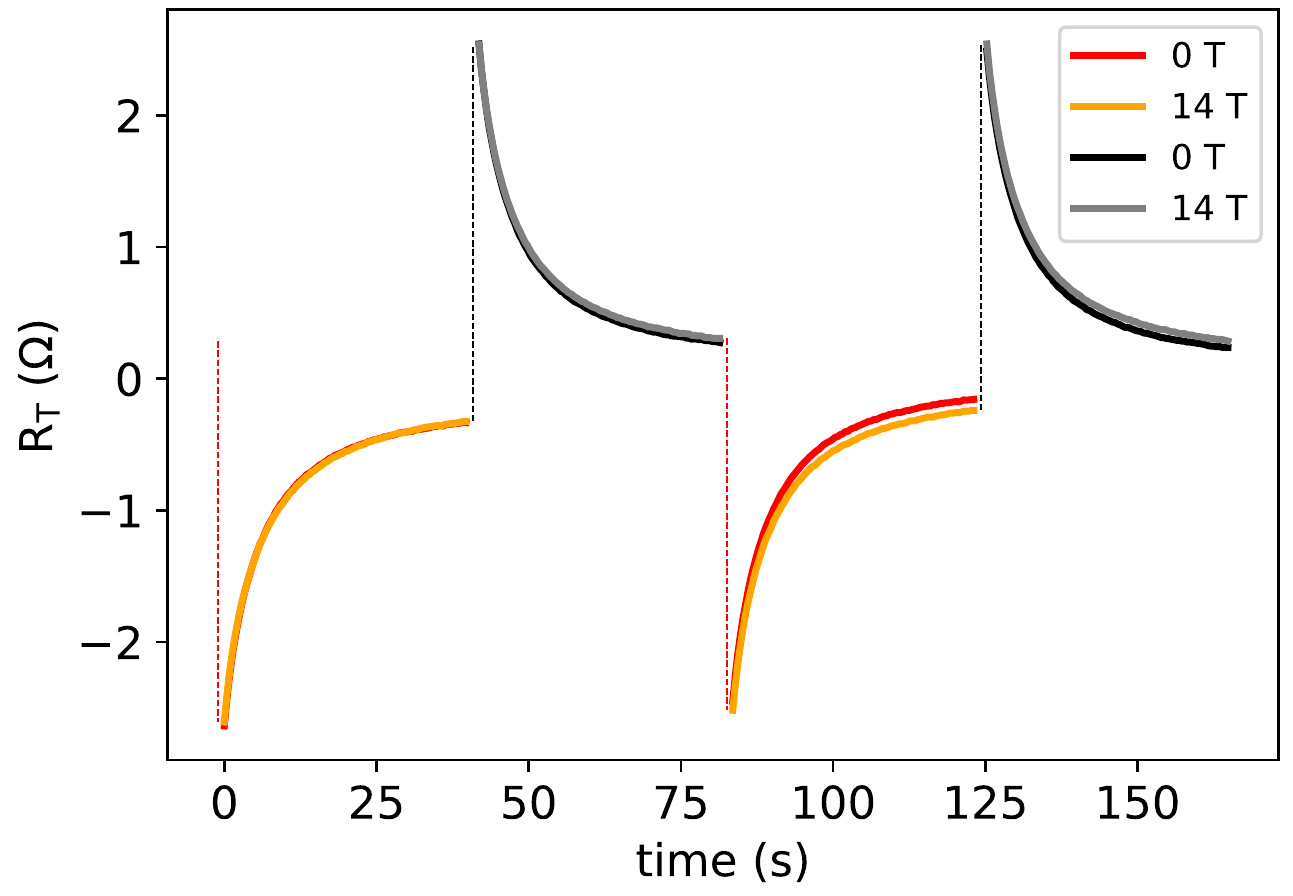}
\end{center}
\caption{Comparison of the switching signal by 100~$\mu$s pulses in the Wheatstone device measured at 0 and 14~T.}
\label{figS15}
\end{figure}
\protect

\end{document}